\documentclass[12pt,preprint]{aastex}
\usepackage{amsmath}
\usepackage{amsfonts}
\usepackage{amssymb}
\usepackage{url}
\usepackage{epsfig}
\usepackage{changepage}
\newcommand\arcpt{${{\lower3pt\hbox{$^{\prime\prime}$}}\atop{\raise4pt\hbox{.}}}$}

\slugcomment{Accepted by the {\it Astronomical Journal}, July 2013}

\shorttitle{Habitable Zones Around Nearby Stars}
\shortauthors{Cantrell et al.}

\begin{document}

\title {The Solar Neighborhood XXIX: The Habitable Real Estate of our
Nearest Stellar Neighbors}

\author {Justin R. Cantrell, Todd J. Henry, Russel J. White}

\affil{Georgia State University, Atlanta, GA 30302-4106}

\email{cantrell@chara.gsu.edu, thenry@chara.gsu.edu,
white@chara.gsu.edu}


\begin{abstract}

We use the sample of known stars and brown dwarfs within 5 pc of the
Sun, supplemented with AFGK stars within 10 pc, to determine which
stellar spectral types provide the most habitable real estate ---
defined to be locations where liquid water could be present on
Earth-like planets.  Stellar temperatures and radii are determined by
fitting model spectra to spatially resolved broad-band photometric
energy distributions for stars in the sample.  Using these values, the
locations of the habitable zones are calculated using an empirical
formula for planetary surface temperature and assuming the condition
of liquid water, called here the empirical habitable zone, or EHZ.
Systems that have dynamically disruptive companions, assuming a 5:1
separation ratios for primary/secondary pairs and either object and a
planet, are considered not habitable.  We use the results of these
calculations to derive a simple formula to predict the location of the
EHZ for main sequence stars based on $V-K$ color. We consider EHZ
widths as more useful measures of the habitable real estate around
stars than areas because multiple planets are not expected to orbit
stars at identical stellar distances. This EHZ provides a qualitative 
guide on where to expect the largest population of planets in the 
habitable zone of main sequence stars. Because of their large numbers
and lower frequency of short-period companions, M stars provide more
EHZ real estate than other spectral types, possessing 36.5\% of the
habitable real estate \textit{en masse}.  K stars are second with
21.5\%, while A, F, and G stars offer 18.5\%, 6.9\% and 16.6\%,
respectively. Our calculations show that three M dwarfs within 10 pc
harbor planets in their EHZs --- GJ 581 may have two planets (d with
$m$sin$i$ = 6.1 $M_\earth$; g with $m$sin$i$ = 3.1 $M_\earth$), GJ 667
C has one (c with $m$sin$i$ = 4.5 $M_\earth$), and GJ 876 has two (b
with $m$sin$i$ = 1.89 $M_{Jup}$ and c with $m$sin$i$ = 0.56
$M_{Jup}$).  If Earth-like planets are as common around low mass stars
as recent Kepler results suggest, M stars are the most likely place to
find Earth-like planets in habitable zones.

\end{abstract}

\section{Introduction}

Early astronomers looked to the sky and saw the Moon as a habitable
world covered in vast oceans, Venus as a swampy marshland enshrouded
in clouds, and Mars with grand canals (Lowell 1895). Not one of these
worlds has maintained its promise of abundant life. Instead, the Solar
System, once thought to be teeming with life, may be barren, although
hope remains for environments under the icy crust of Europa (Marion et
al. 2003), in the tiger stripes of Enceladus (Parkinson et al.~2007),
in water under the Martian surface (Malin \& Edgett 2000), or perhaps
lurking somewhere as yet unidentified.  With the discovery of more
than 700\footnote{(May 2013) \url{http://www.exoplanets.org}}
extrasolar planets since 1989, the real estate market in our Solar
System is no longer the only place we might look for evidence of life
beyond Earth. A likely place to search for planets harboring life is
in the habitable zones of nearby stars.  

The term ``habitable zone'' was first coined by Huang (1959) as a
region around a star where a planet could support life. Since then,
there have been many definitions of habitability, most based on the
presence of liquid water on the surface of a planet.  These planets
could be terrestrial in nature, which are known to exist (Borucki \&
the Kepler Team 2010), or perhaps moons of gas-giant planets, which
are suspected to exist (Weidner \& Horne 2010).  Examples of nearby
stars hosting potentially habitable super-Earths include GJ 581, an M
dwarf with potentially two planets in its habitable zone (Mayor et
al.~2009; Vogt et al.~2010b) and GJ 667C, another M dwarf with one
planet in its habitable zone (Anglada-Escud{\'e} et al.~2012).

Kasting et al.~(1993) did pioneering work in describing habitable
zones (HZs) around main sequence stars.  To approximate the location
of the HZ, they introduced a one-dimensional climate model that yields
the distances from main sequence stars where liquid water would be
present, given an initial assumption of a $CO_2$/$H_2O$/$N_2$
atmosphere and an Earth-sized planet.  They describe the inner
boundary of their model as the point at which the atmosphere becomes
saturated with $H_2O$, causing a loss of water via photolysis and
hydrogen escape; the outer boundary is marked by the formation of
$CO_2$ clouds that cool a planet's surface by increasing its albedo
and lowering its convective lapse rate. They give an equation for the
distance from a star, \textit{D}, of the HZ in AU based on the
incident flux that a planet receives, $L/L_\odot$, the star's
luminosity relative to that of the Sun, and $S_{eff}$, the ratio of
outgoing IR flux to the incoming incident flux at the top of the
planet's atmosphere:

\begin{equation}
D= 1AU \Big(\frac{L/L_\odot}{S_{eff}}\Big)\\
\label{eqno1} 
\end{equation}

Using more explicit
terms, Equation~\ref{eqno1} can be used to show the distance from a
star at which a planet would have a given temperature, based on energy
balance (Kaltenegger et al.~2002).  Equation~\ref{eqno2} can then be
used to show that the equilibrium temperature, $T_P$, of a planet at a
distance, $D$, from its host star is a function of the stellar
effective temperature, $T_{eff}$, stellar radius, $R_\star$, and the
planet's Bond albedo, $A$.  Thus, if the stellar $T_{eff}$ and
$R_\star$ are known, we can calculate the range of distances where a
planet with given albedo\footnote{$A =$ 0.3 for the hypothetical
planets used in this paper.} would have a surface temperature suitable
for liquid water, as described by the planet's temperature, $T_P$:

\begin{equation}
{T_P}={T_{eff}}\Big(\frac{R_\star}{2D}\Big)^{1/2}{(1-A)}^{1/4}
\label{eqno2} 
\end{equation}

In this paper, we apply this equation to the nearby population of
stars whose stellar properties are well known to estimate the total
habitable real estate that they provide.  We use the complete sample
of all stars currently known to be within 5 pc of the Sun from Henry
(2012), and an extended 10 pc sample of AFGK stars, as well as binary
properties, photometric and astrometric data.  We present our derived
$T_{eff}$ and $R_\star$ for each star in the sample and discuss the
methods used to derive each star's empirical habitable zone (EHZ).
For multiple star systems, we assess dynamical stability to eliminate
stellar systems unsuitable for long-term planetary orbits.  Our main
goal is to determine, as a function of spectral type, the cumulative
available EHZ in the solar neighborhood.

\section{Motivations for this Study}

We do not yet have a detailed understanding of the architectures of
all types of planetary systems orbiting various types of stars,
although there has been recent progress for planets in close to stars
via the Kepler mission (e.g., Howard et al.~2012 for planets in
orbital periods less than 50 days around FGKM stars).  In this paper
we assume no bias in the final locations of planets around stars,
including formation and migration (i.e., we assume the distribution 
of planets to be uniform in semi-major axis), to assess the integrated EHZs of
various types of stars found in the solar neighborhood. We focus on
the presence of the {\it first} habitable planet around a given star,
although wider EHZs may of course include more than one planet.  This
is an important assessment to make given the limited telescope time,
funding, and energy of astronomers, so here we focus on the question,
``What set of targets might be most appropriate to observe to improve
the odds of detecting at least one habitable planet?''  This question
is posed in wide-ranging arenas, from conversations between students
and faculty when developing research projects, to discussions of
programmatic directions such as those of NASA's Exoplanet Program
Analysis Group and the NASA/NSF Exoplanet Task Force.

In this paper we evaluate stars within 10 pc in a consistent fashion
to assess what spectral types offer the greatest promise for nearby
habitable planets.  The nearby sample contains stars that have been
characterized carefully, and as a population provide the most accurate
snapshot of the stellar content of the Galaxy.  Thus, this sample
provides the foundation for a realistic understanding of the relative
merits of examining different types of stars.  Our results, coupled
with the planet frequency statistics from the Kepler mission, (e.g.,
Borucki and Koch 2011, Howard et al.~2012, and Dressing and
Charbonneau 2013), can provide statistical measures for the of number
of habitable planets within larger stellar populations, and
particularly among volume-limited samples of nearby stars, as we build
to a comprehensive sample of the nearest stars (e.g., Henry et
al.~2006; Henry 2012).  For example, Howard et al.~(2012) show that
the number of super-Earth size planets increases with decreasing
T$_{eff}$ for orbital periods less than 50 days.  Dressing and
Charbonneau (2013) show that for cool stars (T$_{eff}$ $<$ 4000K) the
occurrence rate for planets with 0.5--4 R$_{\earth}$ is 0.9 planets
per star for orbital periods less than 50 days.  They also calculate
that the lower limit on the occurrence rate of Earth-size planets in
the HZ of cool stars is 0.04 with a 95\% confidence.  Using the
population we describe here, this implies that there could be two
Earth-size planets within the EHZs of M dwarfs within 5 pc of the Sun.
Assuming a constant density of M dwarfs to 10 pc (not all of which
have yet been identified) the number of Earth-size planets jumps to
16.  One of the primary motivations of this study is to determine, in
aggregate, how the odds of finding such planets around M dwarfs
compares to other spectral types.  

The results of our habitable real estate calculations, as outlined in
this paper, are particularly valuable in highlighting that the
ubiquitous M dwarfs provide many locales meeting the canonical
definition for habitability.  Current searches for habitable worlds
orbiting the nearest stars have yielded many Jovian planets and a few
terrestrial worlds, but most of the searches to date have been carried
out using the radial velocity technique and have focused on bright FGK
stars that provide many photons to spectrographs (and which are in the
sweet spot for Kepler).  For transit searches, M dwarfs provide higher
contrast ratios for a given planet, whereas for radial velocity and
astrometric searches they provide larger wobbles for a planet than do
more massive FGK stars.  Knowing that the M dwarfs, as a group, are
important stars to search for habitable worlds helps direct our focus
back to the solar neighborhood, in which three-quarters of all stars
are red dwarfs.  Because of their proximity, these stars hold great
promise for the detailed characterization of exoplanets.

\section{Properties of our Nearest Neighbors}

Primary science goals of the Research Consortium on Nearby Stars
(RECONS\footnote{\url{http://recons.org}}) and the Center for High
Angular Resolution Astronomy
(CHARA\footnote{\url{http://www.chara.gsu.edu/CHARA/}}), are to find
and determine the fundamental properties of our nearest stellar
neighbors.  This has led to the most complete assessment to date of
stars within 5 pc of the Sun.

\subsection{The 5 pc Sample}
\label{5pc}

The modern sample of all known stars and brown dwarfs within 5 pc of
the Sun (Henry 2010, 2012), listed in Table~\ref{tab:parx} (assembled
photometry in Table~\ref{tab:phot}) was first published in the
\textit{Observer's Handbook 2010}.  The sample, which is updated
yearly, was created using the combination of several ground-based and
space-based parallax programs, including the General Catalogue of
Trigonometric Stellar Parallaxes (van Altena et al. 1995), Hipparcos
(Perryman et al.~1997), RECONS (Henry et al. 1997, 2006; Deacon et
al.~2005; Jao et al.~2005; Costa et al. 2005), HST (Benedict et
al.~1999, 2002), Gatewood (1989, 1994) and Gatewood et al.~(1992,
1993).  To be included in the sample, a system must have a weighted
mean trigonometric parallax measurement of 200 mas or greater with an
error of 10 mas or less. To create this sample the parallaxes compiled
were combined and weighted based on the individual measurement errors.
The spectral type, with reference, for each star or star system is
given with the astrometric data, including RA, Dec, the weighted mean
trigonometric parallax, and the number of parallaxes included in the
weighted mean, in Table~\ref{tab:parx}.  The closing date for the
sample is 2012.0.

The 5 pc sample contains 67 stars, including the Sun, and 4 presumed
brown dwarfs (spectral types L or T) in 50 systems.  Of the 50
systems, 34 are single, 11 are binaries, and 5 are triples, giving a
multiplicity fraction of 32\%.  This fraction is consistent with
previous volume limited surveys (35\%; Reid \& Gizis 1997).  The
majority (85\%) of the sample are main sequence stars; the exceptions
include Procyon (GJ 280A), which has a slightly evolved spectral type
of F5IV-V, 5 white dwarfs, and the 4 L/T dwarfs.  The spectral type
breakdown includes 1 A, 1 F, 3 G's (including the Sun), 7 K's, 50 M's,
1 L, 3 T's, and 5 white dwarfs.  We do not include white dwarfs and
brown dwarfs in subsequent calculations of habitable real estate, as
they are objects that are cooling continually, resulting in
unsustainable HZs on long ($\sim$Gyr) timescales.

\subsection{An Estimated 10 pc Sample}
\label{10pcsample}

Due to the sparse population of all but the M stars in the 5 pc
sample, it is difficult to draw meaningful statistics to estimate
which stellar spectral types possess the most habitable real estate.
Therefore, we extend our sample to 10 pc for A, F, G and K stars.
Using the Hipparcos Catalog, which is complete to V=9 (Perryman et
al.~1997), we selected all objects with a parallax greater than 100
mas for inclusion into the sample.  We expect that the extended 10 pc
sample is complete for spectral types A through K, given M$_v$=9.0 for
a M0.0V star (Henry et al.~2006).  However, rather than using spectral
type as a selection criterion, we used a color cutoff of $V-K$ $\le$
3.5 as the dividing line between K and M dwarfs (Kenyon \& Hartmann
1995).  As with the 5 pc sample, weighted mean parallaxes from the
General Catalogue of Trigonometric Stellar Parallaxes, The Hipparcos
Catalog and other sources are listed with the astrometry data for the
extended 10 pc sample in Table~\ref{tab:10pca}, as well as spectral
types and references. Available photometric data are listed in
Table~\ref{tab:10pcphot}.  We note that there are no O type or B type
stars within 10 pc.  Because the M star population is not complete out
to 10 pc (Henry et al.~2006), we approximate the total M star
population by scaling by a factor of eight from the 5 pc sample.  For
clarity, we refer to the additional AFGK stars from 5-10 pc as the
``extended 10 pc sample'' and our estimates of all AFGKM stars within
10 pc as the ``estimated 10 pc sample''. In total, the stellar
population of the estimated 10 pc sample consists of 66 AFGK stars in
57 systems.  Broken down by spectral type, the sample contains 4 A
stars, 6 F stars, 21 G stars, 35 K stars and an estimated 400 M stars
within 10 pc.

\subsection{Photometry and Energy Distributions}
\label{phot}

The $UBVRI$ photometry used in this paper, listed in
Tables~\ref{tab:phot} \& \ref{tab:10pcphot}, was extracted from
sources in the literature with preference given to large surveys and
measurements consistent with photometry obtained as part of the CTIO
Parallax (CTIOPI) program, which uses the Johnson-Kron-Cousins
system. $R$ and $I$ magnitudes from the USNO-B1.0 catalog (Monet et
al. 2003) are incorporated in the extended 10 pc sample and both are
rounded to the nearest 0.1 mag.  $R$ magnitudes are averaged from the
first and second epochs.

The majority of $JHK$ photometry is taken from the 2MASS database,
identified as those values with errors listed explicitly in Tables 2
and 4 (Cutri et al.~2003).  In cases of close binaries with magnitude
difference measurements in the literature (e.g., Henry \& McCarthy
1993 and Henry et al.~1999), the optical and 2MASS photometry is used
with the published magnitude differences to split the component fluxes
into individual magnitudes.  Note that stars brighter than $\sim$5 mag
are saturated in 2MASS images and typically have relatively large
photometric errors ($\ge$ 0.2 mag).  Where possible, we use $JHK$
measurements in the literature for these stars (Johnson et al.~1966;
Johnson et al.~1968; Glass 1975; Mould \& Hyland 1976).  These
magnitudes, listed in their unconverted form in Tables~\ref{tab:phot}
\&~\ref{tab:10pcphot} and with specific references listed in the error
columns, were then converted to 2MASS magnitudes using color
transformations from Carpenter (2001).  Of the 16 multiple systems in
the 5 pc sample, eight have spatially unresolved photometry in some or
all passbands and are marked as ``Joint'' in the Notes section of
Table~\ref{tab:phot}. Similarly, 10 of the 19 multiple systems in the
extended 10 pc sample have spatially unresolved photometry in some or
all passbands and are listed as ``Joint'' in the Notes section of
Table~\ref{tab:10pcphot}.

\section{Methodology to Derive the EHZ}

The HZ around a star is primarily a function of the total energy
output of a star that reaches the surface of a planet.  This can be
determined if the stellar temperature and radius are known, which we
calculate for our sample of stars as described in Section~\ref{SED}.
However, a range of other factors, such as atmospheric pressure,
composition, and cloud cover play roles in determining the surface
temperature of a planet. To account for this, in Section~\ref{EHZ} we
follow previous studies and approximate these effects by using
empirical temperature constraints provided by planets in the Solar
System.

\subsection{SED Fitting used to Derive Stellar Temperature and Radius}
\label{SED}

Although spectral types are often used to estimate stellar effective
temperatures, the various methods for determining spectral types are
inhomogeneous and often depend upon the spectral range used.  As an
alternative, we fit synthetic stellar spectra to broad-band energy
distributions to determine effective temperatures and radii. In
particular, we use photometric measurements spanning $UBVRIJHK$ (0.3
to 2.4 $\mu$m) in conjunction with model spectra generated using the
PHOENIX code (Hauschildt et al.~1999). This prescription is only used
for single stars and for stars in multiple systems in the 5 pc and
extended 10 pc samples with at least four spatially resolved
photometric measurements in different filter bandpasses.  Estimates
for unresolved multiples are discussed in Section~\ref{bhz}. 

These available models span the temperature ranges of $T_{eff}$ of 10,000 K to 7000 K
in 200 K increments and 6900 K to 2000 K in 100 K increments. The model
spectra have a resolution of 2 \AA\, and range from 10 \AA\, to
500 $\mu$m.  The models were convolved with $UBVRIJHK$ filter
responses to create synthetic photometry.  For the $UBVRI$ synthetic
photometry, we took zero points from Bessell et al.~(1998) and filter
responses from 
CTIO\footnote{\url{http://www.ctio.noao.edu/instruments/filters/}}. 
$JHK$ filter responses and zero points were from
2MASS\footnote{\url{http://www.ipac.caltech.edu/2mass/releases/allsky/doc/}}.
The zero points adopted are listed in Table \ref{tab:zp}. We
adopt log~$g$ values of 4.5 or 5.0 for all stars.  

Solar metallicity is adopted for all stars except GJ 191 (Fe/H =
$-$0.98; Woolf and Wallerstein 2005) and GJ 451 (Fe/H = $-$1.16; Valenti
and Fisher 2005), for which we adopt metallicities of $-$1.0.  The
assumption of solar metallicity for the remainder of our 10 pc
sample is based on the 36 FGK stars that overlap with Valenti and
Fisher (2005).  These stars have an average metallicity of Fe/H =
$-$0.048 with a standard deviation of 0.168 dex.

\begin{figure}
\centering
\begin{tabular}{cc}
\epsfig{file=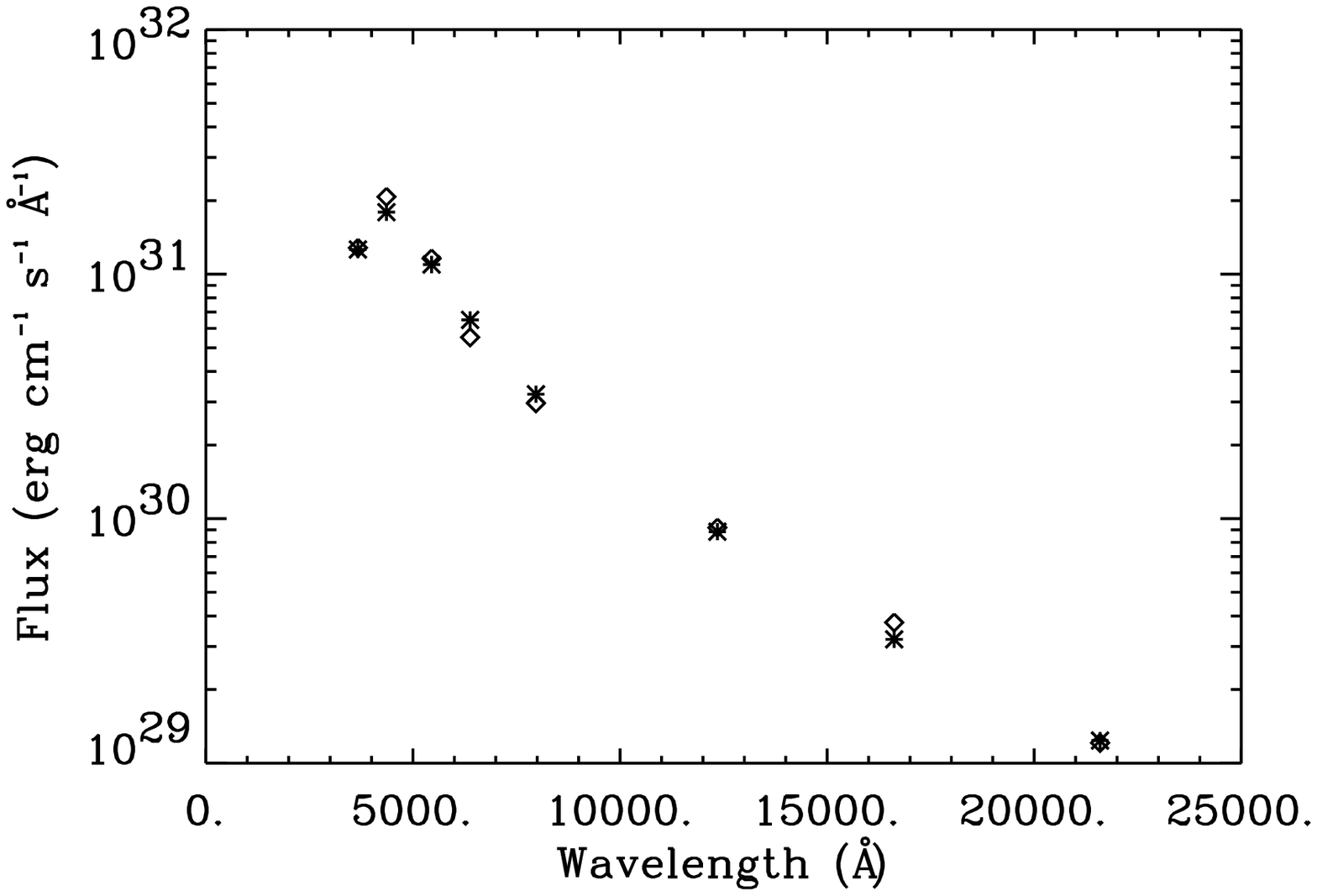,width=0.50\linewidth,clip=} &
\epsfig{file=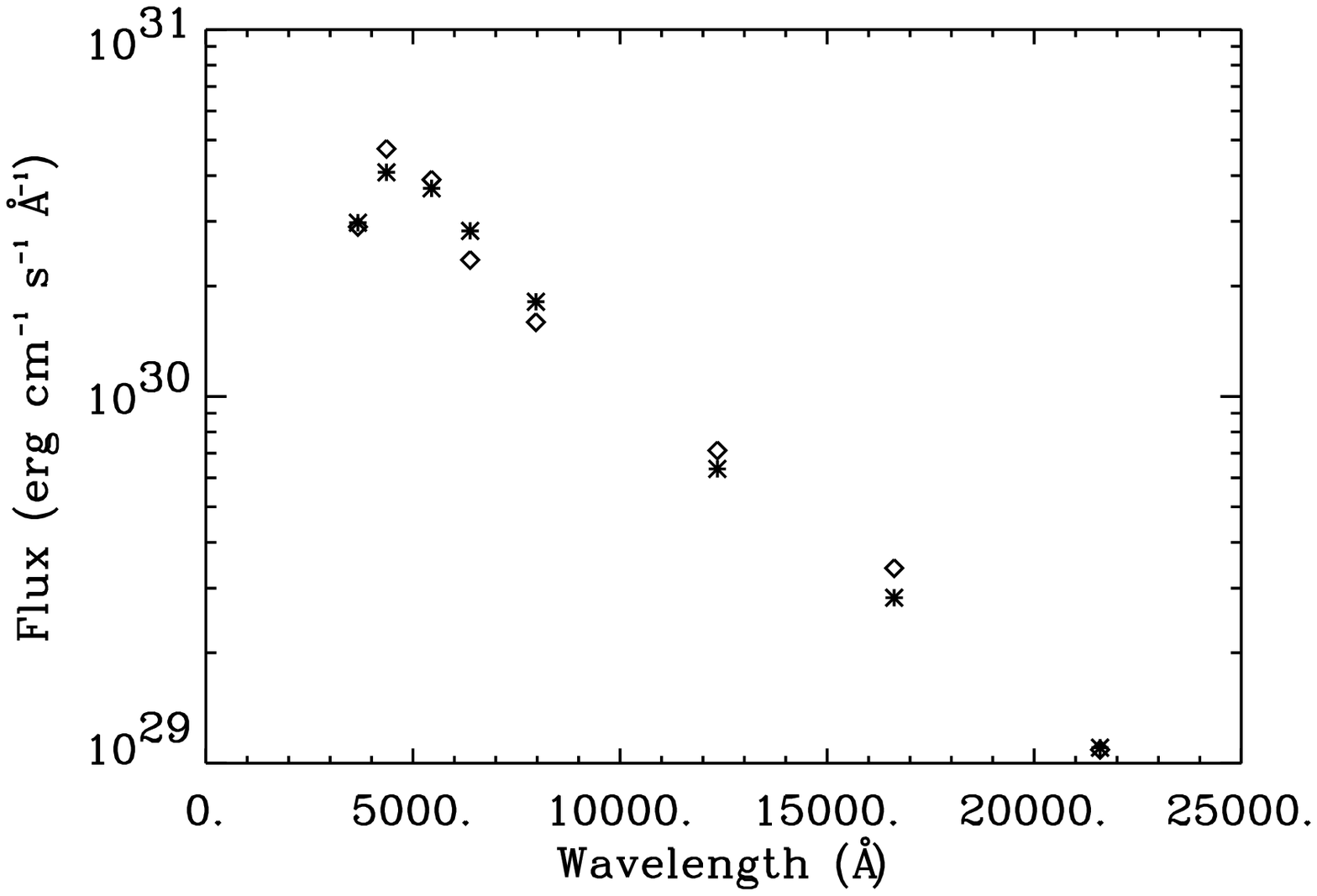,width=0.50\linewidth,clip=} \\
\epsfig{file=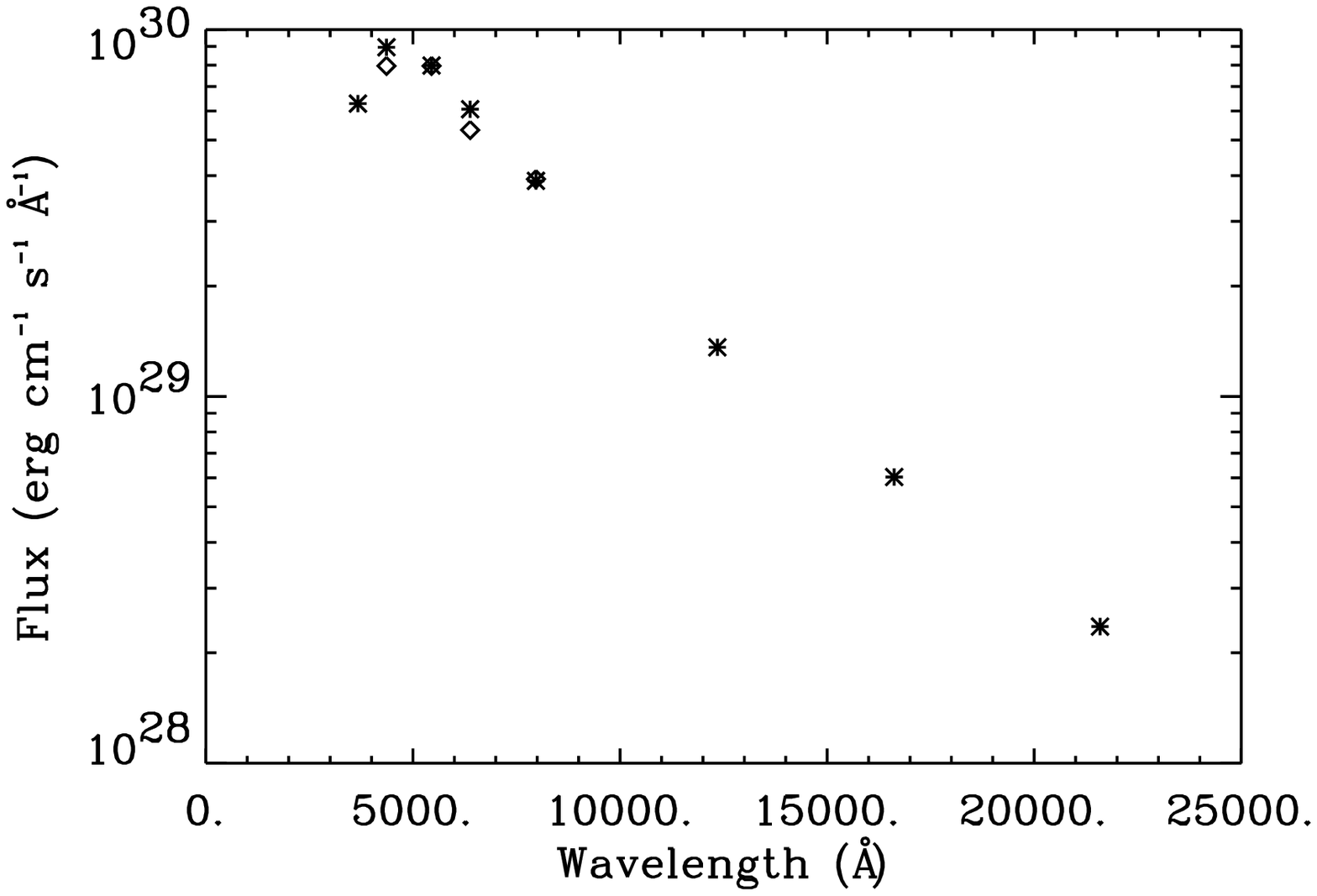,width=0.50\linewidth,clip=} &
\epsfig{file=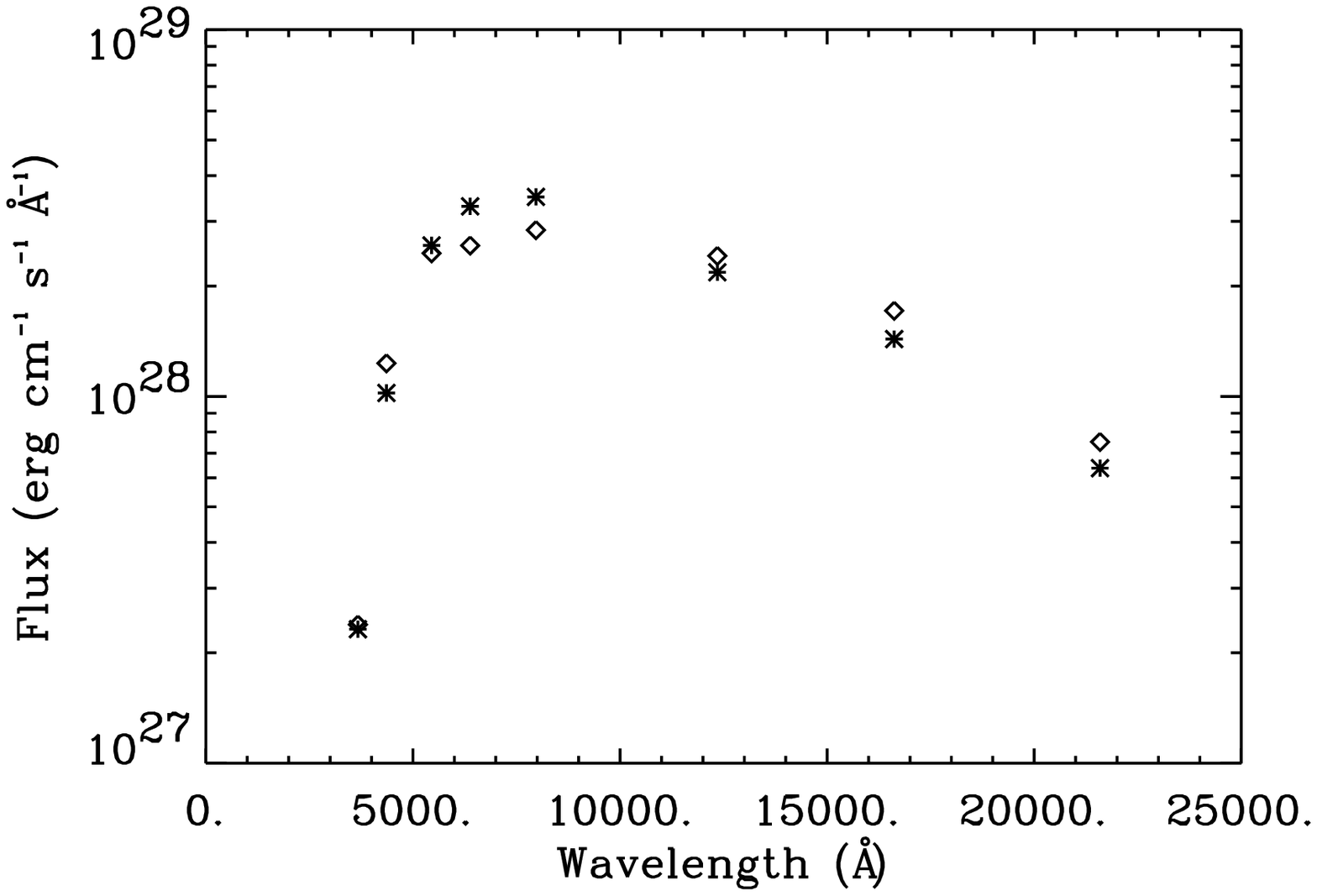,width=0.50\linewidth,clip=} \\
\epsfig{file=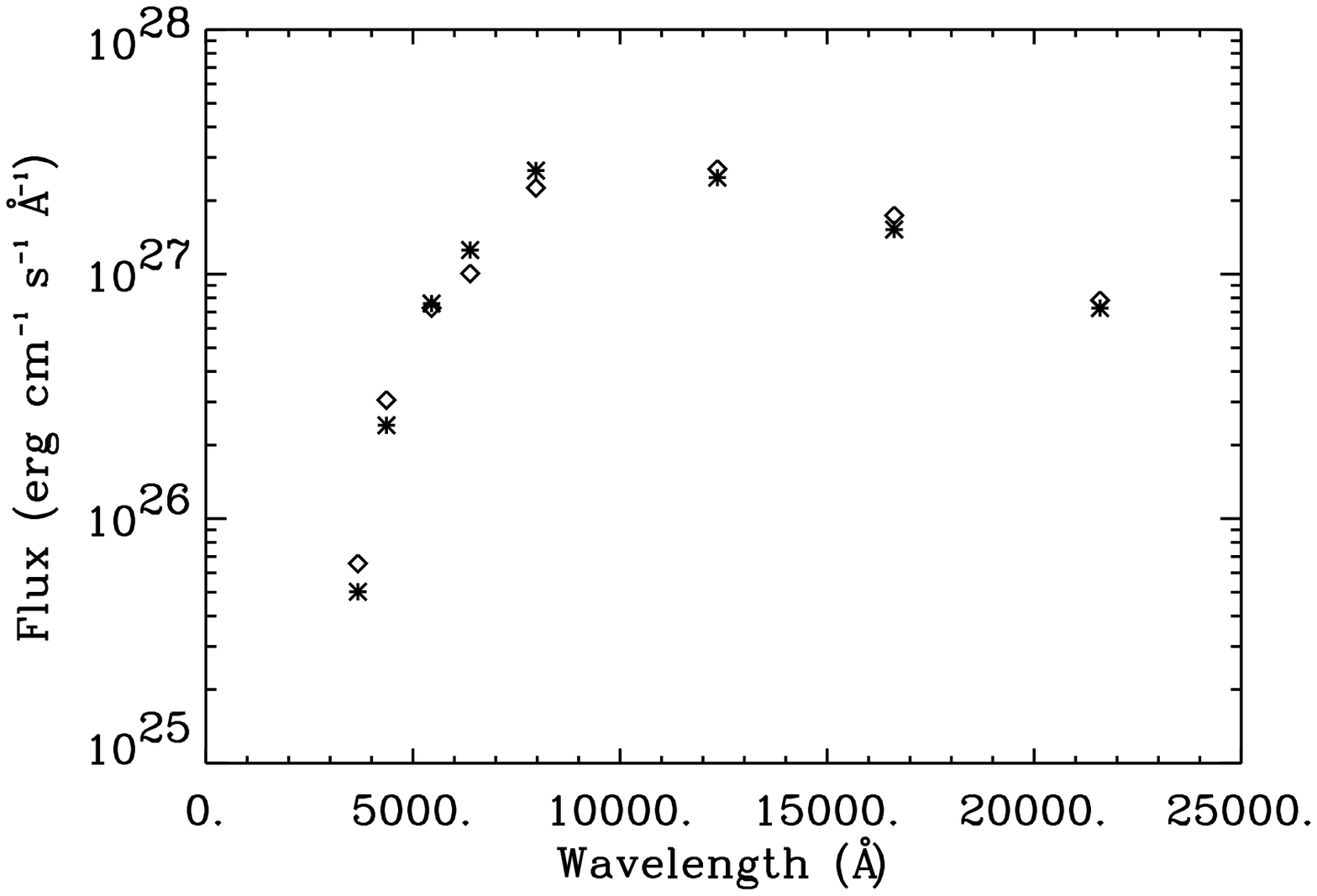,width=0.50\linewidth,clip=} &
\epsfig{file=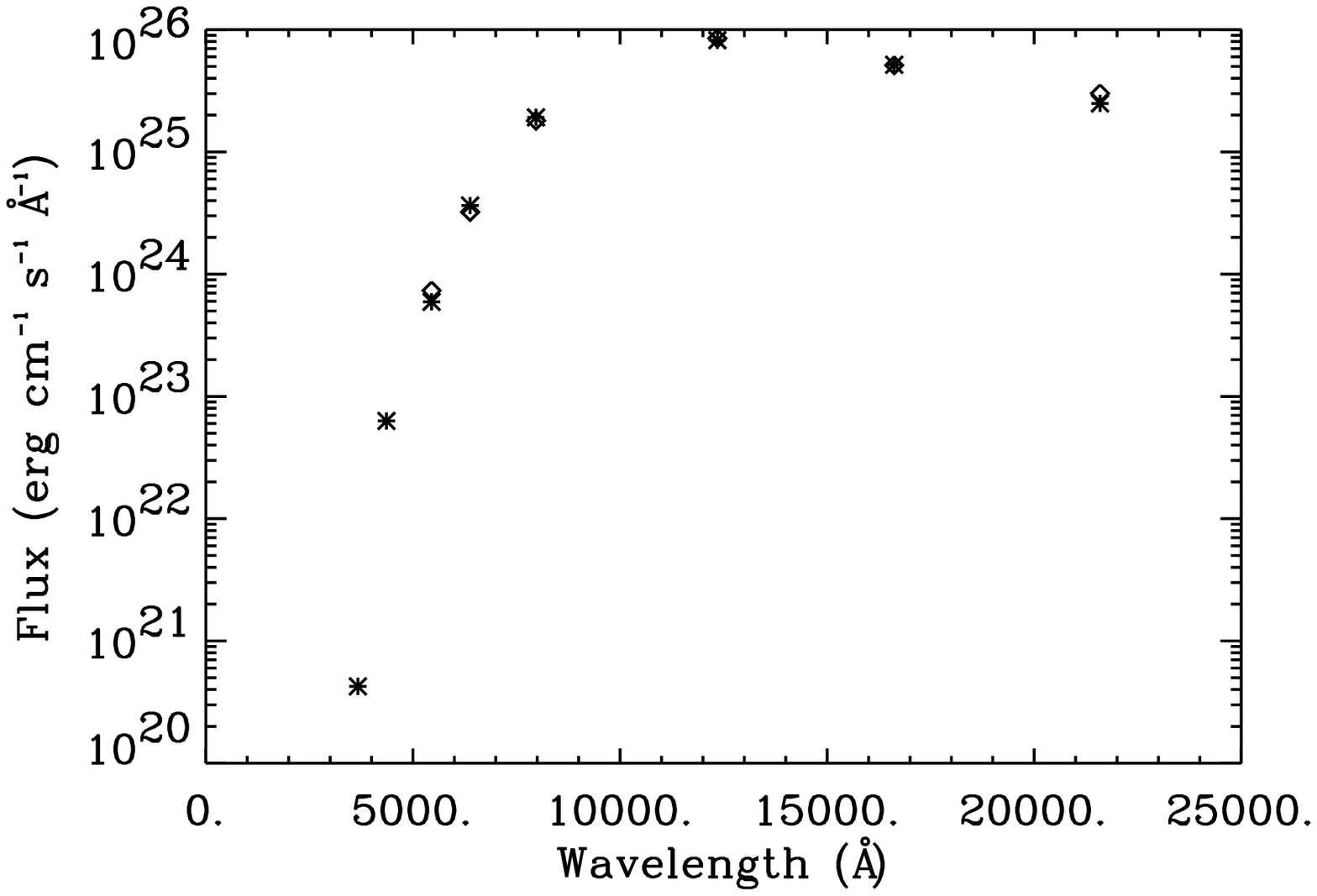,width=0.50\linewidth,clip=} \\
\end{tabular}
\caption{Examples of model ($\ast$) and observed ($\Diamond$) flux
values are shown for (\textit{top-left to right}) GJ 244A (A1.0V), GJ
280A (F5.0IV$-$V), (\textit{middle-left to right}) GJ 599A (G2.0V), GJ
380 (K7.0V), (\textit{bottom-left to right} GJ 273 (M3.5V), SCR
1845-6357A (M8.5V).  In each case the points represent $UBVRIJHK$
photometry.  Model values for stars with incomplete photometry, e.g.,
GJ 559A missing UJHK and SCR 1845-6357A missing UB, are plotted for
completeness.}
\label{fig:fits}
\end{figure}

The flux values in the model spectra are given as a surface flux that
must be scaled by a radius to a known distance for fitting with
observed integrated flux values.  A stellar radius grid from 0.001
R$_\odot$ to 3.00 R$_\odot$ with a step size of 0.001 R$_\odot$ is
calculated for each model spectrum.  Each spectrum and radius
combination is then convolved with filter response and the zero point
data referenced above to derive consequent fluxes observed at Earth.
These integrated fluxes are then fit via a $\chi^2$ minimization
routine, written in IDL and described in Equation~\ref{eqno5}, to
compare the model flux with the observed flux across each filter
bandpass. Here, $O$ is the observed integrated flux from the
photometric measurements, $E$ is the estimated integrated flux from
the model grids, $\nu$ is the number of photometric points (degrees of
freedom), and $\sigma$ is the average error in the photometric
measurement. Examples of fits for AFGKM stars in the 5 pc sample are
shown in Figure~\ref{fig:fits}.

\begin{equation}
\chi_\mathrm{red}^2 = \frac{\chi^2}{\nu} = \frac{1}{\nu} \sum {\frac{(O - E)^2}{\sigma^2}}
\label{eqno5} 
\end{equation}

The output is an effective temperature from the model and a radius
that best fits the measured photometric data.  We assume no
interstellar extinction for our sample, and choose four photometric
points as the minimum number needed to make a fit.  When deriving
radii, all stars in the sample are assumed to be spherical and radiate
isotropically, which is not the case for rapidly rotating stars that
may be oblate and experience gravity darkening (e.g., Altair, see
Monnier et al.~2007; Vega, see Aufdenberg et al.~2006).  This is most
common among stars earlier than mid-F spectral type, as these stars
are fully radiative and consequently do not possess an efficient
rotational braking mechanism (e.g., Wilson 1965).  However, all but
one early type star (i.e., earlier than spectral type F5) within 10 pc
have projected rotational velocities less than 100 km/s, which
correspond to projected oblateness values $\lesssim$2\% (Absil et
al.~2008).  These apparently slow rotating stars include Sirius A (GJ
244, A1V, 16 km/s; Royer 2002), Procyon A (GJ 280, F5V-IV, 4.9 km/s;
Fekel 1997), Vega (GJ 721, A0V, 24 km/s; Royer 2007), and Fomalhaut
(GJ 881, A4V, 93 km/s; Royer et al.~2007).  We note that despite
having a small $v$sin$i$ value, Vega is believed to be rapidly
rotating with a nearly pole on orientation (Aufdenberg et al. 2006).
An additional rapidly rotating star is Altair (GJ 768, A7V, 217 km/s;
Royer 2007) which has an oblateness of 18\% determined from
interferometric measurements by the CHARA Array (Monnier et al. 2007).
As expected, our derived radii and effective temperatures for Altair
and Vega are intermediate to the polar and equatorial values listed in
Monnier et al.~(2007) and Aufdenberg et al.~(2006), respectively.
Because the EHZ boundaries are a function of the square root of the
total luminosity (see Section~\ref{EHZ}), our adopted methodology
should yield reliable estimates of the habitable real estate these
stars provide, even if the stellar temperatures vary with stellar
latitude.

As a consequence of the limited temperature resolution of the PHOENIX
grids used ($\Delta$T=200K from 10000K to 7000K and $\Delta$T=100K
from 6900K to 2000K), the fitting routine can determine a slightly
larger radius coupled with a cooler temperature, or vice
versa. Systematic uncertainties in the PHOENIX models are such that a
$\Delta$T of less than 100K are unreliable (Hauschildt, priv. comm.).
As the Stefan-Boltzmann law shows R$_\star$$^2$T$_{eff}$$^4$$\sim$L$_\star$, the luminosity
determined from the SED remains the same as long as the radii and
effective temperatures move in opposite directions, and therefore
the EHZ, being a function of the square root of the luminosity, does
not change.  Of key importance, the output radii and temperatures
determined here allow us to compare our results directly to those
found via interferometric techniques.

\begin{figure}
\includegraphics{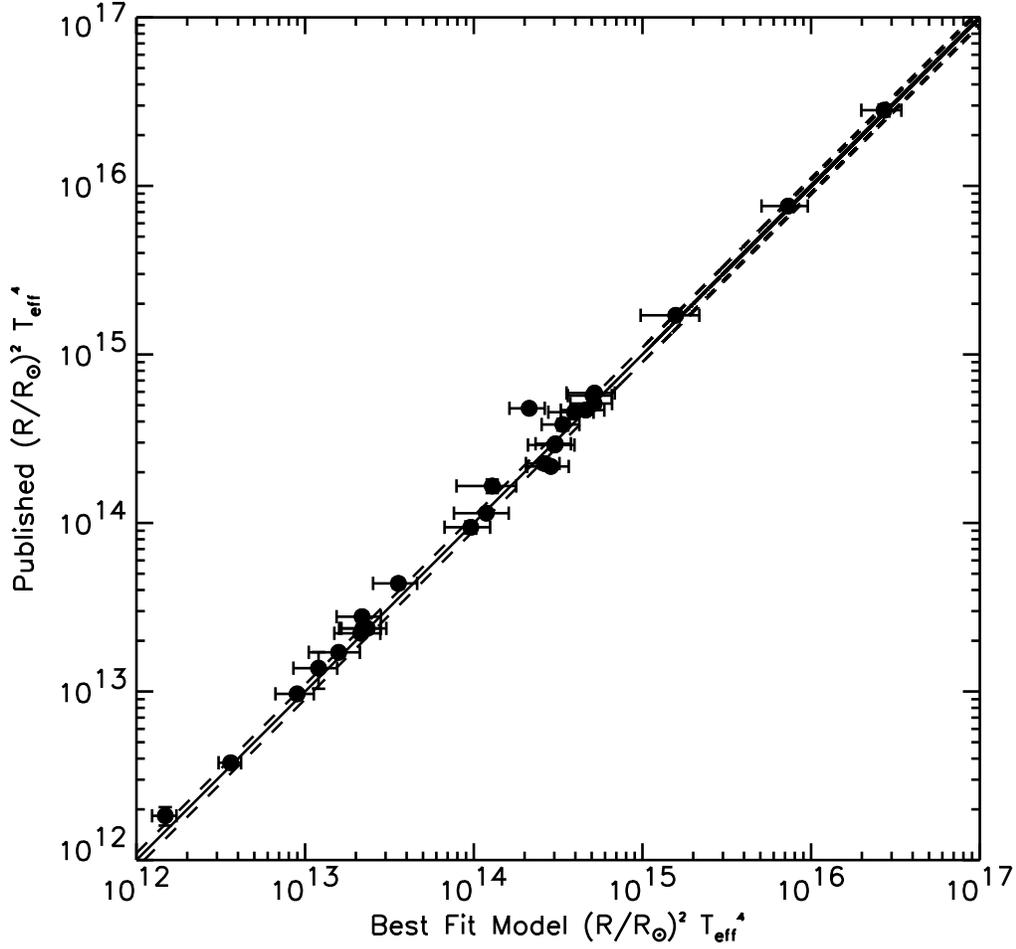}
\caption{Model fits vs published $(R/R_\odot)^{2}T^{4}$ values. The
solid line illustrates 1:1 agreement and the dashed lines represent
10\% offsets. Error bars are 1$\sigma$ for the model and given errors
for the interferometrically derived values used to derive the
published values.}
\label{fig:lumz}
\end{figure}

Of the stars investigated, 28 have spatially resolved angular
diameters from long baseline interferometric instruments such as
CHARA, PTI, and VLTI (Table~\ref{tab:temp}).  Fifteen of these
measurements are from the recent effort of Boyajian et al.~(2012).
Because these measurements determine sizes to within a few percent, we
use them to test the accuracy of the radii determined in our fitting
process. The published radii are on average 7.4\% larger than our
derived radii, which shows a systematic effect.  Our derived model
$T_{eff}$ are on average 2.4\% hotter than derived from
interferometric measurements.  A comparison of our model versus
published values is shown in Figure~\ref{fig:lumz}.  The temperature
uncertainties correspond to spectral type uncertainties of 1-2
spectral subclasses, similar to the error associated with most
spectral classification methods.  As the average values indicate, the
overprediction of temperature leads to the expected under-prediction
of radius, and the combination yields a more accurate luminosity, and thus
consistent EHZ. Given this agreement, we adopt our model values for
T$_{eff}$ and R/R$_\odot$ in final calculations of the habitable real
estate.


A few stars are worthy of note.  Procyon is slightly evolved
(F5.0IV-V) and has a highly constrained log~$g$ of 4.05$\pm$0.04
(Fuhrmann et al.~1997) measured via high resolution spectroscopy in
concert with masses determined using astrometry from the Procyon-white
dwarf orbit.  We adjusted the log~$g$ to 4.0 for this star and
recomputed the $T_{eff}$ and $R$.  This, nevertheless, yielded
identical values within the model grid resolution as those of
log~$g$=4.5.  The model spectra used in this work have a low
temperature limit of 2000 K, which is the value derived for the three
intrinsically faintest stars in this sample: SCR 1845-6357A (M8.5V),
DEN 1048-3956 (M8.5V), and LP 944-020 (M9.0V).


\subsection{Habitable Zones of Single Stars}
\label{EHZ}

Kasting et al.~(1993) use a one-dimensional climate model to calculate
HZs around single main sequence stars.  A one-dimensional climate
model characterizes a planet's global temperature by dividing the
planet into latitudinal bands, and treats the planet as uniform with
respect to longitude.  These one-dimensional radiative-convective
models are a good approximation of global temperature, but more
complicated 3-D global climate models are needed to account for the
complex physical interactions associated with oceans, clouds, and land
surface processes.  These inputs add parameters such as land/ocean
surface coverage and clouds, which can vary from planet to planet,
complicating the overall goal to characterize the EHZ of a star.  Here
we use the one-dimensional model to generalize the EHZ, and adopt a
modified version of the one-dimensional model based on the ``Venus and
early Mars criterion'' from Selsis et al.~(2007). In that work, they
argue that empirical evidence shows that Venus has not had water on
its surface for at least one billion years, and Mars had water on its
surface around 4 billion years ago.  The solar fluxes at those times
were 8\% and 28\% lower, respectively (Baraffe et al.~1998).  Venus
(0.72 AU today) and Mars (1.52 AU today) would need to be at distances
of $\sim$0.75 AU and $\sim$1.77 AU, respectively, to receive these
levels of solar flux today.  

Selsis et al. (2007) provide a method for estimating the inner and
outer edges of HZs for stars with T$_{eff}$ = 3700K-7200K.  The
stars in the sample discussed here range in temperatures from 2000K
to 10000K.  We have therefore chosen to derive new relations for the
HZ boundaries that span the entire stellar temperature regime of our
sample, and thereby provide a consistent methodology for all stars
in the sample.

In defining our EHZ inner and outer
boundaries, we assume a planet with an atmosphere, radius, mass, and
Bond albedo (0.3) that matches Earth (Kasting 1996).  This leads to
the EHZ Equations~\ref{eqno3} and~\ref{eqno4}, used to determine the
empirical surface temperature of an Earth-like planet that satisfies
the ``Venus criterion'' (we adopt 0.80 AU on the suggestion of
Kasting, priv. comm.) and ``Mars criterion'' (1.77 AU). The resulting
inner and outer radii of the EHZ correspond to equilibrium
temperatures of 285 K to 195 K, respectively.

\begin{equation}
R_{EHZ inner}=\frac{R_\star}{2}{(0.7)}^{1/2}\Big(\frac{T_{eff}}{285K}\Big)^2
\label{eqno3} 
\end{equation}

\begin{equation}
R_{EHZ outer}=\frac{R_\star}{2}{(0.7)}^{1/2}\Big(\frac{T_{eff}}{195K}\Big)^2
\label{eqno4} 
\end{equation}

Note that these only depend on R$_\star$ and T$_{eff}^2$, or
essentially the square root of the star's luminosity. Values for both
samples are listed in Tables~\ref{tab:hz} and \ref{tab:10pchz}.  Only
stars used in the cumulative EHZ calculations are listed, except for
the cases of Sirius (GJ 244A) and Procyon (GJ 280A), which provide
useful benchmarks.

\subsection{Habitable Zones of Multiple Star Systems}
\label{bhz}

Although stars in multiple systems are often avoided in planet
searches, planets have been found in binary and multiple systems
(e.g. Patience et al 2002; Raghavan et al.~2006; Eggenberger et
al. 2007).  The potentially dynamically disruptive effects of any
close stellar companion must be considered when assessing the
possibility of formation, long-term dynamical stability, and
ultimately the habitability of planets in multiple star systems.  The
$\alpha$ Centauri triple system, at a distance from the Sun of 1.34 pc
for the G2V-K0V pair (GJ 559 AB) and 1.30 pc for the wide M5V
companion (GJ 551), includes the nearest set of Sun-like stars, and
provides a test case to study the habitability of multiple stars.  The
G-K pair has an orbit with a semimajor axis of 17$\farcs$57 and
eccentricity of 0.518 (Pourbaix et al.~2002).  This gives periastron
and apastron distances of 11.33 AU and 26.67 AU, respectively.
Barbieri et al.~(2002) show for $\alpha$ Centauri A that planets can
form in stable orbits on a timescale of 5 Myr.  Using numerical
simulations, they show that not only could planets form, but in some
models they formed directly in the habitable zone.  Quintana et
al. (2007) similarly find that binary separations greater than 10 AU
did not inhibit the formation of terrestrial planets at 2 AU.
Consistent with this, Wiegert \& Holman (1997) show that the orbit of
a planet can be stable as long as the ratio of the semimajor axis of
the binary to that of the planet is more than 5:1.

The 36 multiple systems in the 5 pc and extended 10 pc samples consist
of 27 binaries, 6 triples, and 3 quadruple systems (GJ 423, GJ 570,
and GJ 695), for a total of 84 stars/brown dwarfs.  However, nine are
not main sequence stars, and excluded because of their evolutionary
states (e.g., sub-giants, white dwarfs, and brown
dwarfs). Additionally, 16 are M star companions in the extended 10 pc
AFGK sample.  Because the habitable real estate of M stars out to 10
pc is estimated by scaling the 5 pc results, we do not consider these
M stars in our EHZ calculations. This leaves 59 stars in multiple
systems with potential habitable zones.  For clarity, we emphasize
that the EHZ of each star in a system is calculated separately.

Of the 59 stars, 43 (73\%) have at least 4 spatially resolved
photometric measurements in different filter bandpasses.  For these
stars, the same prescription used in Section \ref{EHZ} is used to
calculate the EHZs. For the 16 stars that are not photometrically
spatially resolved from their nearest companion(s), we estimate the
EHZ locations based on spectral type information for the components.
Using the assembled spectral types listed in Tables~\ref{tab:parx} \&
\ref{tab:10pca}, we estimate the components' $V-K$ color using the
spectral type versus $V-K$ colors relations of Kenyon \& Hartmann (1995),
and estimate the location of the EHZs using the relations described in
Section~\ref{vmk}.

\begin{figure}
\includegraphics{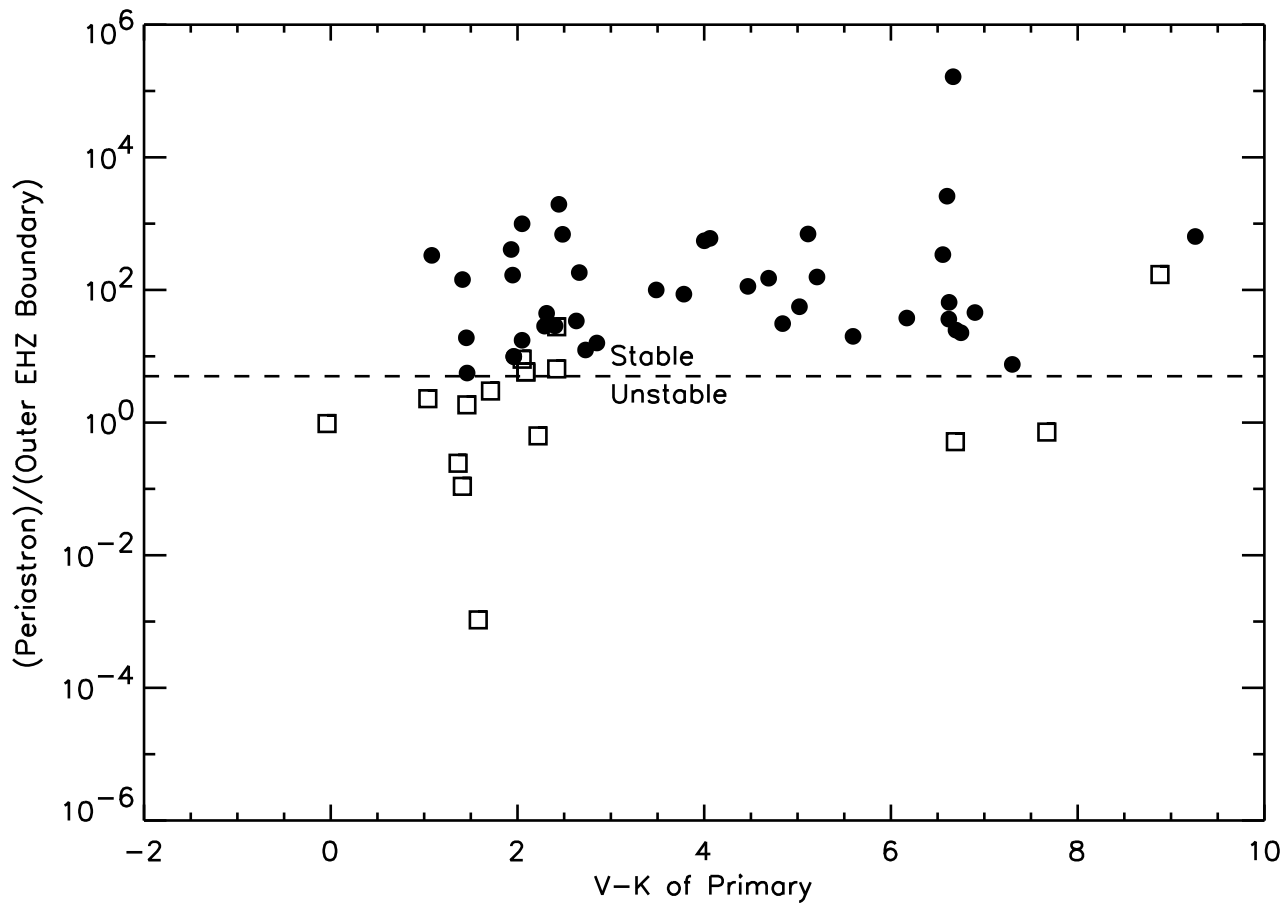}
\caption{The closest approach of a companion star to the outer radius
of the EHZ is plotted versus the primary's $V$-$K$ value.
Photometrically unresolved multiples are plotted as \textit{open
squares}, while resolved components are plotted as \textit{filled
circles}.  Stars with companions that get closer than 5 times the
outer EHZ boundary (\textit{dashed line}) are considered dynamically
unstable planet hosts, and are not included in the total habitable
real estate calculations. GJ 663AB and GJ 663BA are plotted as the
same point as their $\Delta$V and $\Delta$K = 0.  The open square 
at $V-K$ = 8.88, ratio = 173, is SCR 1845-6357, an M dwarf with a 
brown dwarf companion in a highly uncertain orbit.}
\label{fig:sep}
\end{figure}

To assess the dynamical stability of any planets in these EHZs, we
compare the locations of the outer EHZ boundaries to the binary
separations listed in Table~\ref{tab:bin}.  We consider a planet to be
dynamically stable if the periastron distance of a star's nearest
companion is greater than five times the outer EHZ boundary. In cases
where the periastron distance can not be calculated (because its full
orbit solution is not known), we use the projected separation.
Fortunately, these exceptions are all large separation multiples, and
thus minor errors in the adopted separation are likely irrelevant for
dynamical stability considerations.  The ratios of periastron
distances to outer EHZ boundaries are illustrated in
Figure~\ref{fig:sep}.  For each star and its nearest companion, the
ratios range from 0.001 to 164107.  49 of the 59 stars have ratios
greater than 5 to their nearest companions, and thus EHZs in which
planets should be in dynamically stable orbits. The remaining 10 stars
(GJ 53A, GJ 222A, GJ 244A, GJ 280A, GJ 423A, GJ 423B, GJ 713A, GJ
713B, GJ 866A, and GJ 866C) are considered to have EHZs in which
planets would not be in dynamically stable orbits, so are excluded in
our estimate of total EHZ real estate.  Seven of these have ratios
less than 1.0, hinting at the possibility of circumbinary planets.
However, the majority of the EHZ outer radii are only a few times the
binary separations, making it unlikely that these systems would have
dynamically stable circumbinary planets in the EHZ.  Nonetheless, it
is worth noting that a few massive planets have been reported in
circumbinary orbits (Lee et al.~2009; Qian et al 2010).


\section{Discussion}

The described HZ calculations are used to assess the total habitable
real estate in the solar neighborhood and to determine the amount of
habitable real estate as a function of spectral type. Evolved stars,
brown dwarfs, and multiple stars with separations detrimental to the
orbital stability of a planet in the EHZ, as described above, are
excluded in these assessments.  In the 5 pc sample, stars removed from
the analysis due to close companions include GJ 244A, GJ 244B, GJ
280A, GJ 280B, GJ 866A, and GJ 866C.  The distribution by spectral
type, after the removal of these stars in the 5 pc sample is as
follows: 0 A, 0 F, 3 G, 7 K, and 48 M stars.  Similarly, stars removed
from the extended 10 pc sample analysis due to close companions
include GJ 53A, GJ 53B, GJ 222A, GJ 222B, GJ 423A, GJ 423 B, GJ 423C,
GJ 423D, GJ 713A, and GJ 713B.  Stars with evolved spectral types such
as GJ 150, GJ 695A, and GJ 780 are also removed from subsequent
calculations.  By spectral type, the total stellar samples considered
in the final EHZ assessment are 3 A, 4 F, 14 G, 34 K, and an estimated
384 M stars.

\subsection{The EHZ ``Width''}

Using our initial assumptions of a terrestrial ``Earth-like'' planet
as the basis for our EHZ, we estimate the habitable real estate using
linear AU separations from the central star, essentially the width of
the EHZ.  In Table~\ref{tab:hzsp} we present the EHZ width totals for
each spectral type in the 5 pc and total 10 pc samples.

The cumulative EHZ width for stars in the 5 pc subsample is 8.8 AU,
including 2.6 AU for the 3 G stars (including the Sun) and 2.9 AU for
the 7 K stars. The 48 M dwarfs in the 5 pc sample \textit{en masse}
provide 3.3 linear AU available for habitable planets, or 38\% of the
available EHZ.  The dominant contribution of M dwarfs to the EHZ width
is demonstrated clearly using the estimated 10 pc sample.  The total
EHZ width for the estimated 10 pc sample is 71.5 AU, including 13.2 AU
for the 3 A, 4.9 AU for the 4 F stars, 11.9 AU for the 14 G stars, and
15.4 AU for the 34 K stars.  The estimated 384 M stars \textit{en
masse} provide 26.1 AU of linear EHZ.  This accounts for 36.5\% of the
total EHZ. Thus, by spectral type, M stars \textit{en masse} provide
the largest EHZ real estate.

\subsection{Predicting the Size of the Habitable Zone from $V-K$ Colors}
\label{vmk}

Given the rapid pace of exoplanet discovery, it would be helpful to
have a tool to easily and accurately predict the location of the EHZ
to determine whether or not a planet resides within it. Predicting the
EHZ of a star based on spectral classification can be problematic due
to the inhomogeneity in classification and spectral types being
determined over different wavelength ranges.  As shown in Henry et
al.~(2006), the $V-K$ color is a useful temperature diagnostic for the
A through M stars that dominate the solar neighborhood.  This relation
can also be very helpful in determining a rough estimate of the EHZ
based on observable photometry, and may be easily scaled to larger
populations.

\begin{figure}
\includegraphics{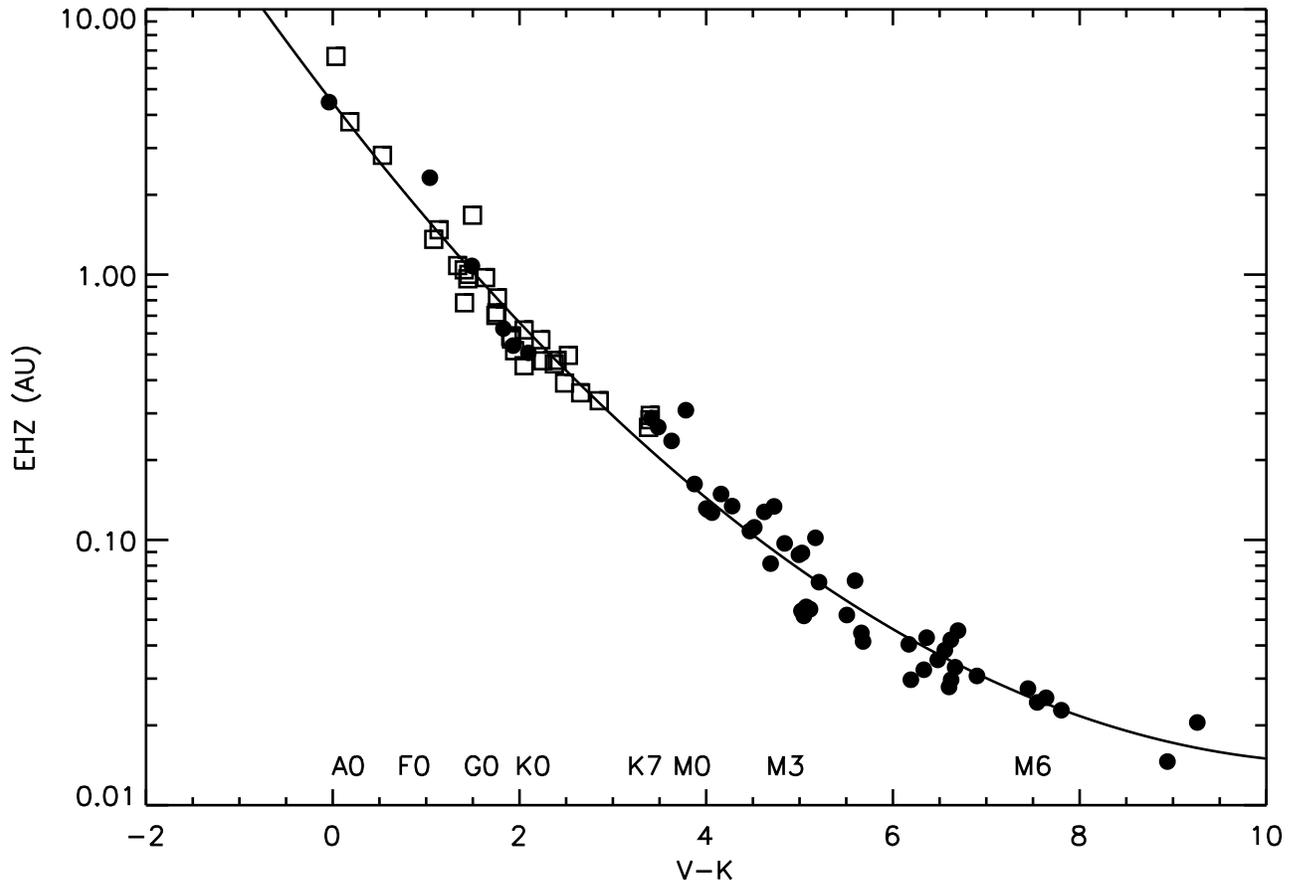}
\caption{Empirical habitable zone (EHZ) widths for the 5 pc
(\textit{filled circles}) and extended 10 pc samples (\textit{open
squares}). The best fit relation (Equation~\ref{eqvmk} in text) is
overplotted.}
\label{fig:10pchz}
\end{figure}

We use the results from the 5 pc and extended 10 pc samples to derive
a relation between $V-K$ color and the size and location of the EHZ.
As in the computation of total habitable real estate, we removed
binary stars with unresolved photometry, as well as stars known to be
evolved.  We do, however, use stars such as Sirius (GJ 244) and
Procyon (GJ 280) for which EHZs have been determined, even though
their companions corrupt their EHZs.  Their white dwarf companions do
not significantly contribute to their luminosities or $V-K$, and their
inclusion improves the statistics of our fit.  We fit a second order
polynomial, described by Equation~\ref{eqvmk}, to the $V-K$ colors and
computed EHZ widths shown in Figure ~\ref{fig:10pchz}.

\begin{equation}
Log(EHZ_\mathrm{width (AU)}) = 0.648 - 0.457 (V-K) + 0.021(V-K)^2
\label{eqvmk} 
\end{equation}

Using the same method, a relationship for $V-K$ color and the inner and
outer radii of the EHZs are described by Equations~\ref{eqvmkin} and
\ref{eqvmkout}, respectively.  These relations are only valid for main
sequence stars.

\begin{equation}
Log(EHZ_\mathrm{inner (AU)}) = 0.593 - 0.457(V-K) + 0.021(V-K)^2
\label{eqvmkin} 
\end{equation}

\begin{equation}
Log(EHZ_\mathrm{outer (AU)}) = 0.922 - 0.457(V-K) + 0.021(V-K)^2
\label{eqvmkout} 
\end{equation}

As a check on these relations, we use these to calculate the total EHZ
width by spectral type of the estimated 10 pc sample and compare these
values to the direct calculations described in Sections~\ref{EHZ} \&
\ref{bhz}.  We find the total EHZ widths from the empirical relations
differ on average from the calculated total widths by $-$12.7\%,
$-$6.4\%, 1.6\%, 4.7\%, and 0.3\% for A, F, G, K, and M stars
respectively; negative values correspond to underpredictions of the
EHZ width totals.  The higher percentage differences for A type and F
type totals are due to the relatively small populations within 10 pc
and the effects of large 2MASS photometric errors due to brightness.  The results
imply that this relation is useful for quickly estimating the amount
of habitable real estate for a population with known $V-K$ values.  We
also test how well the predicted inner and outer boundaries from our
relations agree for any given star.  On average, these predictions
yield values consistent to 3\% with a dispersion of 22\% for both
inner and outer boundaries for AFGKM stars in our samples.  These
dispersions can be interpreted as the uncertainty in the locations of
these boundaries from these relations.

\subsection{Planets in the EHZs of Nearby Stars}

Of the 14 confirmed planetary systems within 10 pc of the Sun for
which orbits have been determined, five contain multiple planets (GJ
139, GJ 506, GJ 581, GJ 667C, and GJ 876).  All 14 systems are listed
in Table~\ref{tab:exop} with published values for semimajor axis and
eccentricity, as well as calculated inner and outer radii of the EHZ
from this work.  Three of the systems, GJ 581, GJ 667C, and GJ 876
have planets in the EHZ.

GJ 581, an M2.5V star at a distance of 6.25 pc, has six proposed
planets, but the existence of GJ 581g and GJ 581f are currently
debated (see: Vogt et al.~2010; Andrae et al.~2010; Gregory 2011;
Anglada-Escud{\'e} 2010; Tuomi 2011).  If real, GJ 581g, with a
semimajor axis of 0.146 AU, orbits in the EHZ in a presumed circular
orbit.  Using the CHARA Array, von Braun et al.~(2011) recently
measured the size of the star and derived HZ boundaries of R$_{in}$=
0.11 AU and R$_{out}$= 0.21 AU.  Our EHZ is somewhat closer in to the
star, R$_{in}$= 0.083 AU and R$_{out}$= 0.179 AU, but still places GJ
581g in the EHZ.  The differences are due to the our calculated
luminosity (L= 0.11$L_\odot$) being 8\% lower than von Braun et
al.~(2011), as they adopted an extinction of A$_V$= 0.174 for GJ 581,
which they note as unexpected for a star at this distance.  GJ 581d,
with semimajor axis of 0.22 AU and eccentricity of 0.38, also moves in
and out of the EHZ of GJ 581.

GJ 667C, an M1.5V star at a distance of 7.23 pc, hosts two planets,
and possibly four (Anglada-Escud{\'e} et al.~2012).  Although GJ 667Cb
does not lie within the EHZ, GJ 667Cc ($m$~sin~$i$ of 4.5 $M_\earth$)
lies within the EHZ for the majority of its orbit.  With a semimajor
axis of 0.123 AU and eccentricity $<$0.27, it may lie completely
within the EHZ once the eccentricity is more highly constrained.

\begin{figure}
\includegraphics{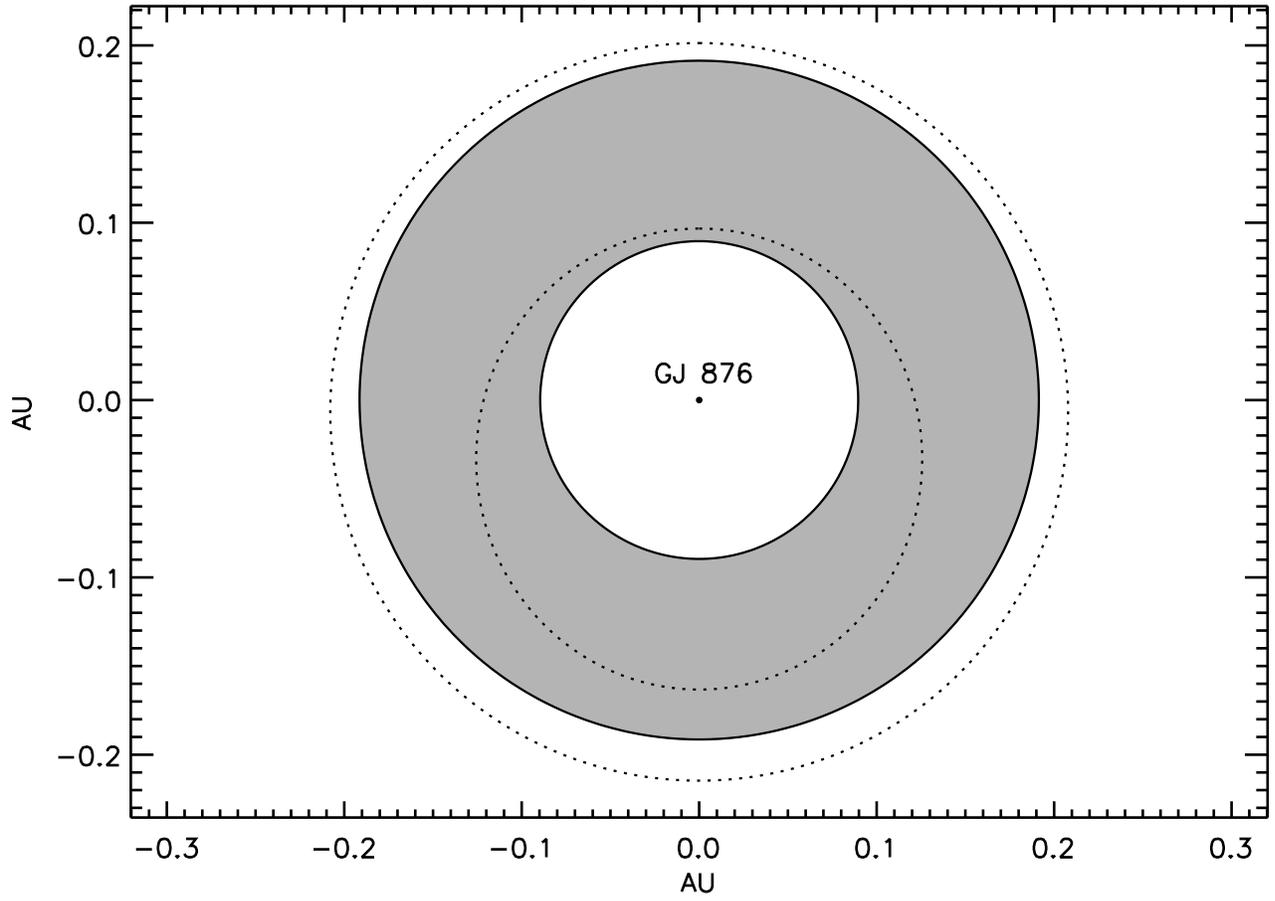}
\caption{The EHZ in this figure is shown as a shaded disk, along with
the orbits of the two planets (dotted circles) in the GJ 876 system
nearest the EHZ.}
\label{fig:gj876}
\end{figure}

GJ 876, an M3.5V star at a distance of 4.66 pc, hosts four planets
(Rivera et al. 2010).  Our calculations show an EHZ spanning
0.090-0.191 AU.  Rivera et al.~(2010) report orbital fits for two
planets near the EHZ with semimajor axes of 0.13 AU (GJ 876c) and 0.21
AU (GJ 876b), and eccentricities of 0.25 and 0.03, respectively.  As
shown in Figure~\ref{fig:gj876}, this allows for GJ 876c to be in the
EHZ of its host star for the full duration of its orbit, while GJ 876b
lies just outside the EHZ. Although these planets are not considered
terrestrial ($m$~sin~$i$ of 0.56 and 1.89 $M_J$), the possibility
exists that they could have terrestrial-like moons that could be
habitable.

\subsection{Complications to Habitability}

The previous sections provide estimates of EHZs based on the
requirement of liquid water on a planetary surface.  Of course, a
planet's location in the EHZ of a host star does not guarantee its
habitability.  A host of other factors, such as planet size,
atmosphere, magnetic fields, and even plate tectonics, play vital
roles in determining the habitability of a planet in the EHZ.  Without
a sufficiently thick atmosphere, biologically harmful radiation can
penetrate to the surface of a planet.  On Earth, atmospheric $CO_2$
levels are kept in check by the carbon-silicate cycle.  Known to
regulate climate temperatures through a negative feedback, this cycle
allows for as much as 60 bars (Kasting 1996) of $CO_2$ to be locked
away in rock and sediments.  This slow carbon-silicate cycle requires
that water be present, because without water, the atmospheric $CO_2$
cannot be sequestered as carbonate.  A magnetosphere on Earth also
plays an important role by deflecting harmful charged particles.
Tectonic activity may be one of the key factors in keeping a planet
habitable (Doyle et al.~1998).  Without water, the lithosphere of a
planet may become a stagnant lid, halting tectonic activity and the
sequestration of $CO_2$.

There is a temporal constraint on habitability as well, as the HZ may
change considerably over the lifetime of a star (Kasting et al.~1993;
Kasting 1996; Tarter et al.~2007; Selsis et al.~2007).  Putting these
complications aside, the requirement of liquid water on the surface of
a planet is a good first order approximation to habitability.

\section{Summary}

We assess the sample of stars currently known to be within 5 pc of the
Sun for the purpose of determining the habitable real estate and its
dependence on spectral type.  Because of the sparse population of high
mass stars within 5 pc of the Sun, we expand this sample for AFGK
stars to 10 pc; there are no O or B stars within 10 pc of the Sun.
After eliminating evolved stars, substellar objects, and close
multiples in which planets in HZs would be dynamically unstable, we
use the final sample to estimate the EHZs for stars in both the 5 pc
and extended 10 pc samples.  We do not consider circumbinary habitable
zones in this work, but EHZs are calculated in the same fashion as
single stars for each of the 49 components in multiple systems that
satisfy our dynamical stability constraint.

Using PHOENIX models convolved with filter response curves, we fit
observed $UBVRIJHK$ photometry for each object, assuming spherical,
non-rapidly rotating stars with solar metallicity and log~$g$ values
of 4.0 to 5.0, with the 2 exceptions being the metal poor stars GJ 191 and GJ 451. 
This fitting process allows us to determine a radius
and $T_{eff}$ for each star that is then used to determine its
surrounding EHZ, calculated using a modified ``Venus and early Mars
criterion'' from Selsis et al.~(2007).

We use estimates of linear AU to map the EHZ of each star and sum by
spectral type \textit{en masse}: 48 M dwarf stars used in the 5 pc
sample provide more habitable real estate (3.3 AU) than the three G
dwarfs stars (2.6 AU) and seven K stars (2.9 AU) found within 5 pc of
the Sun.  Even after extending the sample of AFGK stars to 10 pc, the
anticipated sample of M dwarfs within 10 pc (not all have yet been
identified) possess more EHZ real estate than any other spectral type,
spanning $\sim$26 AU compared to 13.2 AU, 11.9 AU, and 15.4 AU for
each of the A, G, and K types (the F stars provide only 4.9 linear AU
of EHZ). The result is a natural consequence of the large relative
numbers of M dwarfs, and the frequency of close companions that
declines with mass.

As a population M dwarfs provide more options and more habitable real
estate than their more massive counterparts.  Furthermore, recent
results from Kepler show that for stars with T$_{eff}$>4000K within 5
pc of the Sun, there is likely to be at least 2 Earth-size planets in
the HZ.  That number increases to 16 within 10 pc (Dressing \&
Charbonneau 2013).

Using the 5 pc and extended 10 pc samples, we derive relations between
$V-K$ color and EHZ width and inner and outer limits. Comparisons of
color predicted locations suggest they are comparable to uncertainties
associated with habitability assumptions (e.g., Section
\ref{EHZ}). Thus, these color relations are practical tools for
estimating the EHZs of stars using commonly available photometric
measurements. The relations for the inner and outer radii of the EHZ
are helpful for quickly estimating whether or not a known planet or
disk is within the EHZ.  The relation for EHZ size is useful in
predicting the habitable real estate available in a stellar
population.  In particular, we consider the results for the 14
extrasolar planetary systems known within 10 pc of the Sun.  The three
systems with planets in the calculated EHZs --- GJ 581, GJ 667C, and
GJ 876 --- are all M dwarfs.  In total, as many as four planets
circling these stars spend at least part of their time in the EHZs,
providing an ideal set of targets for future efforts to detect
biosignatures.

\subsection{Acknowledgments}

The authors would like to thank J.~Kasting for extremely helpful
advice and guidance.  We would also like to thank the RECONS team for
the generous amount of data and support provided. Data used in this
paper were acquired via support from the National Science Foundation
(grants AST 05-07711 and AST 09-08402) and through the continuous
cooperation of the SMARTS Consortium. This project was funded in part 
by NSF/AAG grant no. 0908018.  This research has made use of
the SIMBAD database, operated at CDS, Strasbourg, France, data from
the Two Micron All Sky Survey, which is a joint project of the
University of Massachusetts and IPAC, funded by NASA and NSF, and the
Sixth Catalog of Orbits of Visual Binary Stars, operated by the US
Naval Observatory.

\clearpage


\clearpage

\begin{deluxetable}{lrccccccc}
\rotate
\tabletypesize{\tiny}
\tablecaption{Five Parsec Sample}
\tablewidth{0pt} 
\tablehead{
           \colhead{Name}&
           \colhead{LHS}&
           \colhead{RA 2000.0}&
           \colhead{DEC}&
           \colhead{$\pi$}&
           \colhead{err}&
           \colhead{$\#$ of $\pi$}&
           \colhead{SpType}&
           \colhead{Ref}}
\startdata
 Sun             &\nodata&  \nodata     & \nodata      & \nodata  &\nodata  &\nodata& G2.0V      & \nodata \\
 GJ    1         &      1&   00 05 24.4 &  $-$37 21 27 &   .23032 &  .00090 &  2   &  M1.5V      &  Hen10  \\
 GJ 1002         &      2&   00 06 43.2 &  $-$07 32 17 &   .21300 &  .00360 &  1   &  M5.0V      &  Hen94  \\
 GJ   15 A       &      3&   00 18 22.9 &  $+$44 01 23 &   .27987 &  .00060 &  3   &  M1.5V      &  Hen94  \\
 GJ   15 B       &      4&   00 18 22.9 &  $+$44 01 23 &   .27987 &  .00060 &  3   &  M3.5V      &  Hen94  \\
 GJ   35         &      7&   00 49 09.9 &  $+$05 23 19 &   .23270 &  .00181 &  2   & DZ7         &  YPC95  \\
 GJ   54.1       &    138&   01 12 30.6 &  $-$16 59 57 &   .26908 &  .00299 &  2   &  M4.5V      &  Hen94  \\
 GJ   65 A       &      9&   01 39 01.3 &  $-$17 57 01 &   .37370 &  .00270 &  1   &  M5.5V      &  Hen94  \\
\enddata
\label{tab:parx}
 \tablecomments{\\
      Bil06    Biller et al.~(2006)                               \\
      CNS91    Catalog of Nearby Stars, Gliese \& Jahreiss (1991) \\
      Cos06    Costa et al.~(2006)			          \\
      Gra01    Gray et al.~(2001)                                 \\
      Gra06    Gray et al.~(2006)                                 \\
      Hen94    Henry et al.~(1994)                                \\
      Hen02    Henry et al.~(2002)				  \\
      Hen04    Henry et al.~(2004)                                \\
      Hen06    Henry et al.~(2006)				  \\
      Hen10    Henry et al.~(2010)				  \\
      Jao08    Jao et al.~(2008)                                  \\
      Joh53    Johnson \& Morgan (1953)                           \\
      Joh10    Johnson et al.~(2010)                              \\
      McC04    McCaughrean et al.~(2004)                          \\
      Mon01    Montes et al.~(2001)                               \\
      Rei95    Reid et al.~(1995)                                 \\
      Rei04    Reid et al.~(2004)                                 \\
      Sch07    Schmidt et al.~(2007)                              \\
      Tor06    Torres et al.~(2006)                               \\
      YPC95    Yale Parallax Catalog, van Altena et al.~(1995)    \\
}
\end{deluxetable}	   																		    

\begin{deluxetable}{lrccccccc}
\rotate
\tabletypesize{\tiny}
\tablecaption{Extended Ten Parsec Sample}
\tablewidth{0pt} 
\tablehead{
           \colhead{Name}&
           \colhead{LHS}&
           \colhead{RA 2000.0}&
           \colhead{DEC}&
           \colhead{$\pi$}&
           \colhead{err}&
           \colhead{$\#$ of $\pi$}&
           \colhead{SpType}&
           \colhead{Ref}}
\startdata
GJ 17     &    5      &    00 20 04.2   &  $-$64 52 29     & .11647   &  .00016  &  2  &     F9.5V      &   Gra06  \\
GJ 19     &    6      &    00 25 45.1   &  $-$77 15 15     & .13407   &  .00011  &  2  &     G0.0V      &   Gra06  \\
GJ 33     &    121    &    00 48 23.0   &  $+$05 16 50     & .13426   &  .00049  &  2  &     K2.5V      &   Gra06  \\
GJ 34 A   &    123    &    00 49 06.3   &  $+$57 48 55     & .16823   &  .00046  &  2  &     G3.0V      &   Mon01  \\
GJ 53 A   &    8      &    01 08 16.4   &  $+$54 55 13     & .13267   &  .00074  &  2  &     G5.0V      &   Joh53  \\
GJ 66 A   &   \nodata &    01 39 47.6   &  $-$56 11 47     & .12999   &  .00208  &  2  &     K5.0V      &   Mon01  \\
GJ 66 B   &   \nodata &    01 39 47.6   &  $-$56 11 36     & .12999   &  .00208  &  2  &     K5.0V      &   Mon01  \\
GJ 68     &    1287   &    01 42 29.8   &  $+$20 16 07     & .13275   &  .00049  &  2  &     K1.0V      &   Gra06  \\
\enddata
\tablecomments{
* Designates component to above system with different GJ number.      \\*
Bat89    Batten et al.~(1989)                                         \\*
CNS91    Catalog of Nearby Stars, Gliese \& Jahreiss (1991)	      \\*
Far10    Farrington et al.~(2010)                                      \\*
Gra06   Gray et al.~(2006)                                            \\*
Mon99   Montes et al.~(1999)                                          \\*
Mon01   Montes et al.(2001)					      \\*
Pas94   Pasquini et al.~(1994)					      \\*
Ruc95   Ruck \& Smith (1995)					      \\*
Joh53    Johnson \& Morgan (1953)                                      \\*
}
\label{tab:10pca}
\end{deluxetable}


\begin{deluxetable}{llcccccccccccccccc}
\rotate
\tabletypesize{\tiny}
\tablecaption{Extended Ten Parsec Sample: Photometry}
\tablewidth{0pt} 
\tablehead{
           \colhead{Name}&
           \colhead{U}&
           \colhead{Uref}&          
           \colhead{B}&
           \colhead{Bref}&
           \colhead{V}&
           \colhead{Vref}&
           \colhead{R}&
           \colhead{Rref}&
           \colhead{I}&
           \colhead{Iref}&
           \colhead{J}&
           \colhead{err}&
           \colhead{H}&
           \colhead{err}&
           \colhead{K}&
           \colhead{err}&
           \colhead{Notes}}
\startdata
GJ~17     & 4.82       &  Joh66  &  4.80   &   Bes90  &   4.22  & Bes90  &   3.89     & Bes90     &  3.57    &  Bes90    &  3.17   &  Gla75   &  2.87    &   Gla75   &   2.78     &   Gla75   & \nodata    \\    
GJ~19     & 3.53       &  Joh66  &  3.42   &   Bes90  &   2.80  & Bes90  &   2.45     & Bes90     &  2.12    &  Bes90    &  1.72   &  Gla75   &  1.40    &   Gla75   &   1.32     &   Gla75   & \nodata    \\   
GJ~33     &   \nodata  &  \nodata&  6.60   & Bes90    &   5.72  & Bes90  &   5.21     & Bes90     &  4.76    &  Bes90    &  4.24   &  Joh68   &  3.72    &   Joh68   &   3.48     &   Joh68   & \nodata    \\   
GJ~34 A   &   4.04     &  Joh66  &  4.02   & Joh66    &   3.44  & Joh66  &   3.1      & USNOB     &  2.8     &  USNOB    &  2.109  &  0.570   &  2.086   &   0.504   &   1.988    &   \nodata & \nodata    \\   
GJ~53 AB  &   5.96     &  Joh66  &  5.87   & Joh66    &   5.18  & Joh66  &   4.7      & USNOB     &  4.4     &  USNOB    &  3.86   &  Joh68   &  3.39    &   Joh68   &   3.36     &   Joh68   &  joint     \\    
GJ~66 A   &   \nodata  &  \nodata&  6.69   & Hog00    &   5.80  & Hog00  &   \nodata  & \nodata   &  \nodata &  \nodata  &  4.043  &  0.378   &  \nodata &   \nodata &   3.510    &   0.282   & \nodata    \\    
GJ~66 B   &   \nodata  &  \nodata&  6.84   & Mer86    &   5.96  & Mer86  &   \nodata  & \nodata   &  \nodata &  \nodata  &  3.573  &  \nodata &  \nodata &   \nodata &   3.558    &   0.270   & \nodata    \\    
GJ~68     &   6.57     &  Joh66  &  6.08   & Joh66    &   5.24  & Joh66  &   4.7      & USNOB     &  4.3     &  USNOB    &  3.855  &  0.24    &  3.391   &   0.226   &   3.285    &   0.266   & \nodata    \\
\enddata
\tablecomments{
* Designates component to above system with different GJ number.   \\*
JHK \& err from 2MASS unless otherwise noted.                       \\*
``joint'' indicates unresolved photometry.                \\*
Bes90     Bessell (1990)                         \\*
Bre64     Breckinridge \& Kron (1964)                   \\*
Egg65     Eggen (1965)                                 \\*
Gla75     Glass  (1975)                          \\*
Hog00     H{\o}g et al.~(2000) 			     \\*
Joh53     Johnson \& Morgan (1953)                    \\*
Joh66     Johnson et al.~(1966)                   \\*
Joh68     Johnson et al.~(1968)                       \\*
Lep05     L{\'e}pine \& Shara (2005)                  \\*
Lut71     Lutz (1971)                                 \\*
Mer86     Mermilliod (1986) 			     \\*
Oja84     Oja (1984) 				     \\*
Oja93     Oja (1993) 				     \\*
Tor06     Torres et al.~(2006)                        \\*
USNOB     Monet et al.~(2003)                          \\*
}
\label{tab:10pcphot}.
\end{deluxetable}


\begin{deluxetable}{lrrc}
\tabletypesize{\footnotesize}
\tablecaption{$UBVRIJHK$ Zero Points}
\tablewidth{0pt}
\tablehead{\vspace{-15pt} \\
\colhead{Filter}&
\colhead{Zero Point}&
\colhead{$\lambda_{eff}$(\AA)}&
\colhead{Ref}}
\startdata
U & 4.18e-9  &  3660 & Bessell et al.(1998)   \\
B & 6.32e-9  &  4380 & Bessell et al.(1998)   \\
V & 3.63e-9  &  5450 & Bessell et al.(1998)   \\
R & 2.18e-9  &  6410 & Bessell et al.(1998)   \\
I & 1.13e-9  &  7980 & Bessell et al.(1998)   \\
J & 3.14e-10 & 12350 & Cohen et al.(2003)   \\
H & 1.11e-10 & 16620 & Cohen et al.(2003)   \\
K & 4.29e-11 & 21590 & Cohen et al.(2003)   \\
\enddata
\tablecomments{Zero points have units of ergs cm$^2$ sec$^{-1}$ \AA$^{-1}$.}
\label{tab:zp}
\end{deluxetable}


\begin{deluxetable}{llrcccclc}
\rotate
\tabletypesize{\tiny}
\tablecaption{Comparison of Derived Temperatures and Radii for Nearby Stars Using Our SED Fits and Interferometric Results}
\tablewidth{0pt}
\tablehead{  \\
           \colhead{Star}&
           \colhead{SpType}&
           \colhead{$V$-$K$}&
           \colhead{$T_{\rm eff}$Model}&
           \colhead{$T_{\rm eff}$ModelFixR}&
           \colhead{$T_{\rm eff}$Published}&
           \colhead{${R}/{R_\odot}_{Model}$}&
           \colhead{${R}/{R_\odot}_{Pub}$}&
           \colhead{Ref.}}
\startdata
GJ 244 A &  A1.0V   &$-$0.040  &   10000 &     9600   &   9900$\pm$200 & 1.645   &   1.711$\pm$ 0.013     &  Kervella et al (2003a)     \\
GJ 280 A &  F5.0IV-V&   1.028  &    6700 &     6500   &   6524         & 1.898   &   2.048$\pm$ 0.025     &  Kervella et al (2003b)     \\
GJ 559 A &  G2.0V   &   1.490  &    6000 &     5700   &   5810$\pm$50  & 1.103   &   1.224$\pm$ 0.003     &  Kervella et al (2003b)     \\
GJ 71    &  G8.5V   &   1.696  &    5700 &     5400   &   5264$\pm$100 & 0.708   &   0.816$\pm$ 0.013     &  Di Folco et al.(2004)      \\
GJ 631   &  K2.0V   &   1.890  &    5500 &     5300   &   5337$\pm$41  & 0.710   &   0.759$\pm$ 0.012     &  Boyajian et al.(2012)      \\
GJ 166 A &  K0.5V   &   1.932  &    5200 &     5100   &   5143$\pm$14  & 0.735   &   0.806$\pm$ 0.004     &  Boyajian et al.(2012)      \\
GJ 559 B &  K0.0V   &   1.940  &    5400 &     5100   &   5260$\pm$50  & 0.775   &   0.863$\pm$ 0.005     &  Kervella et al (2003b)     \\
GJ 144   &  K2.0V   &   1.954  &    5200 &     5000   &   5135$\pm$100 & 0.689   &   0.743$\pm$ 0.005     &  Di Folco et al.(2004)      \\
\enddata
\tablecomments{ $T_{\rm eff}$Model refers to temperatures from our
model, while, $T_{\rm eff}$ModelFixR refers to temperatures derived
while holding the radius to values obtained through long baseline
interferometry.\\
* GJ 702B does not have enough resolved photometry for a model solution.}
\label{tab:temp}
\end{deluxetable}


\begin{deluxetable}{lllccccccccc}
\rotate
\tabletypesize{\scriptsize}
\tablecaption{Empirical Habitable Zones (EHZs) for the Five Parsec Sample}
\tablewidth{0pt}
\tablehead{\vspace{-15pt} \\
\colhead{}&     
\colhead{}&     
\colhead{}&
\colhead{Model}& 
\colhead{}& 
\colhead{}&
\colhead{}& 
\colhead{}&
\colhead{Model FixR}& 
\colhead{}& 
\colhead{}&
\colhead{}\\ 
\colhead{Name}& 
\colhead{SpType}&     
\colhead{R }& 
\colhead{$T_{\rm eff}$}& 
\colhead{EHZ Inner R}&
\colhead{EHZ Outer R}&
\colhead{EHZ Lin}& 
\colhead{R }& 
\colhead{$T_{\rm eff}$}& 
\colhead{EHZ Inner R}&
\colhead{EHZ Outer R}&
\colhead{EHZ Lin} \\
\colhead{}& 
\colhead{}&     
\colhead{($R/{R_\odot}$)}& 
\colhead{(K)}& 
\colhead{(AU)}&
\colhead{(AU)}&
\colhead{(AU)}& 
\colhead{($R/{R_\odot}$)}& 
\colhead{(K)}& 
\colhead{(AU)}&
\colhead{(AU)}&
\colhead{(AU)}} 
\startdata
%
Sun              &   G2.0V   &   0.917  &     6000  &   0.789  &   1.686  &   0.897  &    1.00    &      5800   &  0.804   &  1.718     &   0.914     \\
DEN 1048-3956    &   M8.5V   &   0.134  &     2000  &   0.013  &   0.027  &   0.014  &    \nodata &   \nodata   &  \nodata &   \nodata  &   \nodata   \\
GJ    1          &   M1.5V   &   0.349  &     3700  &   0.114  &   0.244  &   0.130  &    \nodata &   \nodata   &  \nodata &   \nodata  &   \nodata   \\
GJ   15    A     &   M1.5V   &   0.323  &     3800  &   0.112  &   0.239  &   0.127  &    0.387   &      3700   &  0.127   & 0.271      & 0.144       \\
GJ   15    B     &   M3.5V   &   0.197  &     3200  &   0.048  &   0.103  &   0.055  &    \nodata &   \nodata   &  \nodata &   \nodata  &   \nodata   \\
GJ   54.1        &   M4.5V   &   0.181  &     2900  &   0.036  &   0.078  &   0.042  &    \nodata &   \nodata   &  \nodata &   \nodata  &   \nodata   \\
GJ   65    A     &   M5.5V   &   0.248  &     2600  &   0.040  &   0.086  &   0.046  &    \nodata &   \nodata   &  \nodata &   \nodata  &   \nodata   \\
GJ   65    B     &   M6.0V   &   0.225  &     2700  &   0.039  &   0.084  &   0.045  &    \nodata &   \nodata   &  \nodata &   \nodata  &   \nodata   \\
\enddata
\tablecomments{This table shows model radius, temperature, inner and
outer HZ radius, and the HZ width in AU in columns 3-7.  Columns 8-12
show radius, temperature, inner and outer HZ radius, and the HZ width
in AU based on the method of holding the stellar radius to the values
obtained through long baseline interferometry (see Table 6).  * Sirius
(GJ 244A) and Procyon (GJ 280A) are included as benchmarks, but are 
not used in final EHZ calculations because of companion white dwarfs.
}
\label{tab:hz}
\end{deluxetable}  


\begin{deluxetable}{lllccccccccc}
\rotate
\tabletypesize{\scriptsize}
\tablecaption{Empirical Habitable Zones (EHZs) for the Extended Ten Parsec Sample}
\tablewidth{0pt}
\tablehead{\vspace{-15pt} \\
\colhead{}&     
\colhead{}&     
\colhead{}&
\colhead{Model}& 
\colhead{}& 
\colhead{}&
\colhead{}& 
\colhead{}&
\colhead{Model FixR}& 
\colhead{}& 
\colhead{}&
\colhead{}\\ 
\colhead{Name}& 
\colhead{SpType}&     
\colhead{R }& 
\colhead{$T_{\rm eff}$}& 
\colhead{EHZ Inner R}&
\colhead{EHZ Outer R}&
\colhead{EHZ Lin}& 
\colhead{R }& 
\colhead{$T_{\rm eff}$}& 
\colhead{EHZ Inner R}&
\colhead{EHZ Outer R}&
\colhead{EHZ Lin} \\
\colhead{}& 
\colhead{}&     
\colhead{($R/{R_\odot}$)}& 
\colhead{(K)}& 
\colhead{(AU)}&
\colhead{(AU)}&
\colhead{(AU)}& 
\colhead{($R/{R_\odot}$)}& 
\colhead{(K)}& 
\colhead{(AU)}&
\colhead{(AU)}&
\colhead{(AU)}} 
\startdata
%
%
GJ  17      &       F9.5V     &       0.989 &     6100   &   0.882     &     1.884     &     1.002   &   \nodata  &    \nodata  &    \nodata  &    \nodata  &    \nodata  \\  
GJ  19      &       G0.0V     &       1.709 &     6000   &   1.474     &     3.149     &     1.675   &   \nodata  &    \nodata  &    \nodata  &    \nodata  &    \nodata  \\  
GJ  33      &       K2.5V     &       0.643 &     5200   &   0.417     &     0.890     &     0.473   &     0.695  &       5000  &   0.416     &      0.889  &     0.473   \\  
GJ  34  A   &       G3.0V     &       0.989 &     6100   &   0.882     &     1.884     &     1.002   &   \nodata  &    \nodata  &    \nodata  &    \nodata  &    \nodata  \\  
GJ  66  A   &       K5.0V     &       0.740 &     5100   &   0.461     &     0.985     &     0.524   &   \nodata  &    \nodata  &    \nodata  &    \nodata  &    \nodata  \\  
GJ  66  B   &       K5.0V     &       0.794 &     4900   &   0.457     &     0.976     &     0.519   &   \nodata  &    \nodata  &    \nodata  &    \nodata  &    \nodata  \\
\enddata
\tablecomments{* Designates wide component to above system with
different GJ number.  \\* This table shows model radius, temperature,
inner and outer HZ radius, and the HZ width in AU in columns 3-7 for
the extended 10 pc sample.\\ HZs for GJ 663A, GJ 663B, GJ 667A, and GJ
667B are estimated using the $V$-$K$ relationship (see
Section~\ref{vmk}).}
\label{tab:10pchz}
\end{deluxetable}  
			

\begin{deluxetable}{lrrrrrrrrc}
\tablecaption{Orbital Properties of Multiple Systems}
\tabletypesize{\scriptsize}
\tablewidth{0pt}
\tablehead{ 
\colhead{Star}&
\colhead{Sep}&
\colhead{a}&
\colhead{\textit{e}}&
\colhead{\textit{i}($^\circ$)}&
\colhead{$\Omega$($^\circ$)}&
\colhead{$\omega$($^\circ$)}&
\colhead{P(yr)}&
\colhead{T}&
\colhead{Ref}}
\startdata
 GJ   15 AB         & 40$\arcsec$ & \nodata         &  \nodata     & \nodata   &   \nodata     & \nodata       & \nodata      & \nodata      &  Eggen (1996)                \\
 GJ   34 AB        &   \nodata   &  11$\farcs$99   &    0.497     &  34.76    &   98.43       &    88.59      &   480        &  1889.6      &  Strand (1969)	          \\
 GJ   53 AB        &   \nodata   &  1$\farcs$01    &     0.561    &  106.8    &   47.3        &   152.7       &   21.75      &  1975.74     &  Drummond et al.~(1995)       \\
 GJ   65 AB         &   \nodata   & 2$\farcs$06     &   0.615      &  127.3    &   150.5       &  285.4        &  26.52       &  1971.88     &  Worley \& Behall (1973)     \\
 GJ   66 AB         &   \nodata   &  7$\farcs$82    &     0.534    &  142.8    &    13.1       &   18.37       &   483.66     &  1813.5      &  van Albada (1957)	          \\
 GJ  105 AC-B       & 165$\arcsec$& \nodata         &  \nodata     & \nodata   &   \nodata     & \nodata       & \nodata      & \nodata      &  Golimowski et al.~(2000)    \\
 GJ  105 AC         &   3$\farcs$3&  \nodata        &  \nodata     & \nodata   &   \nodata     & \nodata       & \nodata      & \nodata      &  Golimowski et al.~(2000)    \\
 GJ  166 A-BC       &  83$\arcsec$&  \nodata        &  \nodata     & \nodata   &   \nodata     & \nodata       & \nodata      & \nodata      &  Baize \& Petit (1989)       \\
\enddata
\tablecomments{*asin(i)}
\label{tab:bin}
\end{deluxetable}



\begin{deluxetable}{lcccc}
\tabletypesize{\footnotesize}

\tablecaption{Total EHZ by Spectral Type.}
\tablewidth{0pt}
\tablehead{\vspace{-15pt} \\
\colhead{}&
\colhead{5 pc Sample}&
\colhead{}&
\colhead{Total 10 pc Sample}&
\colhead{}\\
\colhead{SpType}&
\colhead{\# of stars}&
\colhead{EHZ AU} &
\colhead{\# of stars}&
\colhead{EHZ AU}}
\startdata
A &  0(1)   & \nodata   &    3   (4)  &   13.2    \\
F &  0(1)   & \nodata   &    4   (6)  &    4.9    \\
G &  3      &   2.6     &   14  (21)  &   11.9    \\
K &  7      &   2.9     &   34  (35)  &   15.4    \\
M & 48(50)  &   3.3     &  384*(400)* &   26.1*   \\
\enddata

\tablecomments{The numbers used in the table reflect the number of
stars in each sample for which the EHZ was calculated (and include the
Sun). Photometrically unresolved binaries, evolved stars, and
substellar objects for which EHZs were not calculated are excluded
from these totals. *The totals for the M type population to 10 pc are
a factor of eight greater than the population within 5 pc, estimated
via scaling by the volume ($R^3$). Numbers in parentheses are totals
including photometrically unresolved binaries and evolved stars.}

\label{tab:hzsp}
\end{deluxetable}


\begin{deluxetable}{lcccccc}
\tabletypesize{\footnotesize}
\tablecaption{Exoplanets within Ten Parsecs}
\tablewidth{0pt}
\tablehead{\vspace{-15pt} \\
\colhead{Star}&
\colhead{Planet}&
\colhead{Semimajor Axis}&
\colhead{\textit{e}}&
\colhead{Ref}&
\colhead{HZinner}&
\colhead{HZouter}\\
\colhead{}&
\colhead{}&
\colhead{(AU)}&
\colhead{}&
\colhead{}&
\colhead{(AU)}&
\colhead{(AU)}}
\startdata
        GJ 139    &         b    &        0.1207$\pm$0.0020    &          0.0                    &  Pepe et al.~(2011)                &    0.632           &   1.350              \\
        GJ 139    &         c    &        0.2036$\pm$0.0034    &          0.0                    &  Pepe et al.~(2011)                &    0.632           &   1.350              \\
        GJ 139    &         d    &        0.3499$\pm$0.0059    &          0.0                    &  Pepe et al.~(2011)                &    0.632           &   1.350              \\ 
        GJ 144    &         b    &        3.39$\pm$0.36        &          0.702$\pm$0.039        &  Benedict et al.~(2006)            &    0.439           &   0.938              \\ 
        GJ 176    &         b    &        0.066                &          0                      &  Forveille et al.~(2009)           &    0.136           &   0.290              \\
        GJ 442 A  &         b    &        0.46$\pm$0.04        &          0.34$\pm$0.14          &  Tinney et al.~(2011)              &    0.689           &   1.472              \\
        GJ 506    &         b    &        0.05$\pm$5.0e-6      &          0.12$\pm$0.11          &  Vogt et al.~(2010a)               &    0.689           &   1.472              \\
        GJ 506    &         c	 &        0.2175$\pm$0.0001    &          0.14$\pm$0.06          &  Vogt et al.~(2010a)               &    0.719           &   1.537              \\
\enddata
\tablecomments{Bold rows indicate the planet is in the EHZ for at least part of its orbit.  * Planet detection controversial.}
\label{tab:exop}
\end{deluxetable}

\clearpage

\end{document}